\documentclass[english]{article}
\usepackage[T1]{fontenc}
\usepackage{geometry}
\usepackage{units}
\usepackage{mathtools}
\usepackage{amsmath}
\usepackage{amsthm}
\usepackage{amssymb}
\usepackage{stackrel}
\usepackage{graphicx}
\usepackage{color}
\usepackage{orcidlink}

\makeatletter

\newcommand{\noun}[1]{\textsc{#1}}
\providecommand{\tabularnewline}{\\}

\numberwithin{equation}{section}
\numberwithin{figure}{section}
\theoremstyle{plain}
\newtheorem{thm}{\protect\theoremname}[section]
\theoremstyle{plain}
\newtheorem{cor}[thm]{\protect\corollaryname}
\theoremstyle{remark}
\newtheorem{rem}[thm]{\protect\remarkname}
\newcommand{\rr}{}

\usepackage{bbold}
\usepackage[all]{xy}

\makeatother

\usepackage{babel}
\providecommand{\corollaryname}{Corollary}
\providecommand{\remarkname}{Remark}
\providecommand{\theoremname}{Theorem}

\begin{document}
\title{\noun{Three Dixon-Rosenfeld Planes}}
%\author{}
\maketitle
\
{\rr
\begin{center}
{\large David Chester$^{1\ast \;\orcidlink{0000-0003-4662-8592}}$, Alessio Marrani $^{2\dagger\; \orcidlink{0000-0002-7597-1050}}$,
 Daniele Corradetti$^{3\ddagger \; \orcidlink{0000-0001-8086-0593}}$,
Raymond Aschheim$^{1,4\S \; \orcidlink{0000-0001-6953-8202}}$}
%\\[0pt]
%and {\large Daniele Corradetti$^{3\ddagger \; \orcidlink{0000-0001-8086-0593}}$,}\\[5pt]
\bigskip \bigskip

$^1$ Quantum Gravity Research, Topanga, CA, USA \\[5pt]

$^2$ Centre for Mathematics and Theoretical Physics, \\ University of Hertfordshire, AL10 9AB Hatfield, UK
\\[5pt]

$^3$ Grupo de Física Matemática, Instituto Superior Técnico,\\  Av. Rovisco Pais, 1049-001, Lisboa, Portugal\\[5pt]

$^4$ Gauge Freedom Inc., Los Angeles, CA, USA \\[5pt]

\end{center}

\begin{abstract}
Rosenfeld postulated ``generalized'' projective planes, which exploit a correspondence between rank-one idempotents of Jordan algebras $\mathfrak{J}_3(\mathbb{A})$ and points of projective planes $\mathbb{A}P^2$. The isometry groups of the generalized projective planes (which were later defined rigorously as homogeneous spaces) are entries of the Tits-Freudenthal magic square. Given
recent interest in the Dixon algebra $\mathbb{R}\otimes\mathbb{C}\otimes\mathbb{H}\otimes\mathbb{O}$,
we extend Rosenfeld's approach and present three new coset manifolds.
These \textit{Dixon-Rosenfeld planes} have isometry
algebras that are obtained from Tits' magic formula and
involve all tensorial components of the Dixon algebra. We show that
these are the only three planes obtainable with Tits' formula that preserve
the analogy with Rosenfeld's planes. These non-simple Lie algebras generalize $\mathfrak{f}_{4},\mathfrak{e}_{6},\mathfrak{e}_{7}$
and $\mathfrak{e}_{8}$ for the octonionic plane $\mathbb{O}P^{2}$
and in the octonionic Rosenfeld planes $\left(\mathbb{C}\otimes\mathbb{O}\right)P^{2}$,
$\left(\mathbb{H}\otimes\mathbb{O}\right)P^{2}$ and $\left(\mathbb{O}\otimes\mathbb{O}\right)P^{2}$.
We finally investigate the relationships between the isometry algebras
of the Dixon-Rosenfeld planes and the exceptional Lie algebras.
\\ \\
\textit{Keywords:} Projective geometry, Rosenfeld projective planes, division algebras, Dixon algebra
\end{abstract}

\bigskip \bigskip \bigskip

\noindent $^\ast$ davidc@quantumgravityresearch.org

\noindent $^\dagger$ a.marrani@herts.ac.uk

\noindent $^\ddagger$ danielecorradetti@tecnico.ulisboa.pt

\noindent $^\S$ raymond@quantumgravityresearch.org
}
%%\pacs[JEL Classification]{D8, H51}

%%\pacs[MSC Classification]{35A01, 65L10, 65L12, 65L20, 65L70}

%\date{\today}

\newpage

\tableofcontents

\newpage

\section{\noun{introduction and motivation }}

Recent works by Furey and Hughes \cite{Furey 2014,Furey 2015,Furey Hughes}
along with those by Castro Perelman \cite{Castro 2019b,Castro2019}
{\rr revived interest} in the Dixon algebra $\mathbb{T}$ defined by
Goeffrey Dixon in \cite{Dixon} as the tensor product of the four
normed division algebras, i.e., $\mathbb{R}\otimes\mathbb{C}\otimes\mathbb{H}\otimes\mathbb{O}$,
which can be regarded as $\mathbb{C}\otimes_{\mathbb{R}}\mathbb{H}\otimes_{\mathbb{R}}\mathbb{O}$ or $\mathbb{C}\otimes\mathbb{H}\otimes\mathbb{O}$ without risk of ambiguity. These works are a part of a broader research program that involves the application of division algebras to physics, which
has been active for over 80 years \cite{Conway 1937,Gunyadin 1973,Manogue Dray,Huerta 1,Todorov 2018,Masi Nature}.

Although some algebraic features of the Dixon algebra have already been investigated, to our knowledge, no geometrical space was concretely
linked
to $\mathbb{T}$. Moreover, in a wider sense, the research on this
field is mainly organized in two major trends. One is focused on
the Lie groups arising from the magic square over Hurwitz algebras
(see, for example, the recent excellent work of Wilson, Dray and Manogue \cite{Dray Man E8 construction}, following \cite{Dray-Huerta-Kincaid}). Another trend focuses on the study of non-Hurwitz algebras such {\rr as} the
sedenions \cite{Masi Nature} or the Dixon algebra. The present work
unites both these trends, presenting a geometrical construction that
arises from a magic-square approach involving the Dixon algebra. Indeed,
in this work, we show the only three possible Lie algebras obtainable
through the use of Tits' magic formula for $n=3$ over tensorial components
of the Dixon algebra. We then define the only three compact coset
manifolds of real dimension 128, which we call \emph{Dixon-Rosenfeld
planes}, whose isometry group arises from those Lie algebras.

These geometrical constructions might be useful for physical applications of the Dixon algebra. In the usual quantum-mechanical setup, Hermitian matrices represent
physical observables, while the Jordan product over them is obtained by
a plain linearization of the square product. Since the
Dixon algebra $\mathbb{C}\otimes\mathbb{H}\otimes\mathbb{O}$ is not alternative, it is not possible to define a Jordan
algebra over its Hermitian three-by-three matrices. Nevertheless,
it is well known that in the case of alternative composition algebra
$\mathbb{A}$, Hermitian matrices $H_{3}\left(\mathbb{A}\right)$
that are rank-1 idempotent elements of the Jordan algebra, which correspond to points of the projective plane $\mathbb{A}P^{2}$ \cite{McCrim}.
Similarly, the group $\text{Aut}\left(H_{3}\left(\mathbb{A}\right)\right)$
corresponds to the group of isometries of the plane, i.e.,
$\text{Isom}\left(\mathbb{A}P^{2}\right)$. Therefore, the study of
the isometries of a Dixon-Rosenfeld plane, i.e., the Lie algebras we
present in Section 5 and the isometry groups we present in Section
6, might give for, physical purposes, a geometrical equivalent for
a class of Hermitian matrices and their automorphisms, when the algebraic
setup breaks down.

To fully grasp the motivation of this work, we need to expand
the previous statement starting from the geometric correspondence \cite{Corr Notes Octo}
between rank-one idempotents of Jordan algebra $\mathfrak{J}_{3}\left(\mathbb{A}\right)$
over a Hurwitz algebras $\mathbb{A}\in\left\{ \mathbb{R},\mathbb{C},\mathbb{H},\mathbb{O}\right\} $
and elements of the projective planes $\mathbb{A}P^{2}$ that was
introduced by Jordan, von Neumann, and Wigner \cite{Jordan} in 1934.
This correspondence was explored by Tits, Freudenthal and Rosenfeld
in a series of seminal works \cite{Tits,Freud 1965,Rosenfeld-1993,Vinberg}
that led to a geometric interpretation of Lie algebras and to the
construction of the Tits-Freudenthal magic square, i.e.,
\begin{center}
\begin{tabular}{|c|c|c|c|c|}
\hline
$\mathbb{A}/\mathbb{B}$ & $\mathbb{R}$ & $\mathbb{C}$ & $\mathbb{H}$ & $\mathbb{O}$\tabularnewline
\hline
$\mathbb{R}$ & $A_{1}$ & $A_{2}$ & $C_{3}$ & $F_{4}$\tabularnewline
\hline
$\mathbb{C}$ & $A_{2}$ & $A_{2}\times A_{2}$ & $A_{5}$ & $E_{6}$\tabularnewline
\hline
$\mathbb{H}$ & $C_{3}$ & $A_{5}$ & $D_{6}$ & $E_{7}$\tabularnewline
\hline
$\mathbb{O}$ & $F_{4}$ & $E_{6}$ & $E_{7}$ & $E_{8}$\tabularnewline
\hline
\end{tabular}
\par\end{center}

While Freudenthal interpreted the entries of the magic square as different
forms of automorphisms of the projective plane such as isometries,
collineations, homography etc., Rosenfeld investigated the idea that
every row of the magic square is the Lie algebra of the isometry group
of a (generalized) projective plane of tensorial product over Hurwitz
algebras \cite{Rosenfeld-1993}. In Rosenfeld's picture, $F_{4}$
is the isometry group of the \emph{octonionic plane} $\mathbb{O}P^{2}$,
$E_{6}$ is the isometry group of the \emph{bioctonionic plane} $\mathbb{\left(C\otimes\mathbb{O}\right)}P^{2}$,
$E_{7}$ would be that of a generalised projective plane named \emph{quateroctonionic
plane} $\mathbb{\left(H\otimes\mathbb{O}\right)}P^{2}$ and $E_{8}$
would be that of the \emph{octoctonionic plane} $\mathbb{\left(O\otimes\mathbb{O}\right)}P^{2}$
(cf.~\cite{Atiyah Bernd,Baez,Lands-Maniv} for further insights).
It was soon realized that $\mathbb{\left(H\otimes\mathbb{O}\right)}P^{2}$
and $\mathbb{\left(O\otimes\mathbb{O}\right)}P^{2}$ could not be
defined following the same procedure used for projective spaces, and
that they could not satisfy the axioms of projective geometry. Nevertheless,
Rosenfeld's approach was fruitful, since it is possible to define
homogeneous spaces of the correct dimensions having isometry groups
of the desired form and that are therefore dubbed as $\mathbb{\left(C\otimes\mathbb{O}\right)}P^{2}$,
$\mathbb{\left(H\otimes\mathbb{O}\right)}P^{2}$ and $\mathbb{\left(O\otimes\mathbb{O}\right)}P^{2}$
and named \emph{Rosenfeld planes}. In this spirit, in Section 5, we
show three different Lie algebras arising from Tits' formula and related
to the Dixon algebra. In Section 6, we define three Dixon-Rosenfeld
planes as three coset manifolds with 128 real dimensions whose
isometry groups are the obtained Lie algebras.

From the previous discussion, it {\rr follows} that outside the purely algebraical
and geometrical motivations, the goal of this work is to offer a way
of studying a surrogate of the automorphism group of a Jordan algebra
over three-by-three Hermitian matrices that, in the case of coefficients
in the Dixon algebra $\mathbb{C}\otimes\mathbb{H}\otimes\mathbb{O}$,
is not possible (as it is shown in Section 3). More specifically,
we {\rr identified} three different Lie algebras, i.e., $\mathfrak{A}_{1}$,
$\mathfrak{A}_{2}^{\text{enh}}$ and $\mathfrak{A}_{3}$. These arise
from Tits' formula, are related to the isometry groups of the
Dixon-Rosenfeld planes, and replicate the role that $\mathfrak{f}_{4},\mathfrak{e}_{6},\mathfrak{e}_{7}$
and $\mathfrak{e}_{8}$ have in the case of the octonionic plane $\mathbb{O}P^{2}$
and in the octonionic Rosenfeld planes $\left(\mathbb{C}\otimes\mathbb{O}\right)P^{2}$,
$\left(\mathbb{H}\otimes\mathbb{O}\right)P^{2}$ and $\left(\mathbb{O}\otimes\mathbb{O}\right)P^{2}$.
Relationships between $\mathfrak{A}_{1}$, $\mathfrak{A}_{2}^{\text{enh}}$
and $\mathfrak{A}_{3}$ and the exceptional Lie algebras are investigated in Section 7.

\section{\noun{jordan algebras over tensor products}}

In this work, we make use of Tits' formula \cite{BS} to obtain
a Lie algebra in the vector space given by
\begin{equation}
\mathcal{L}_{3}\left(\mathbb{A},\mathbb{B}\right)=\mathfrak{der}\left(\mathbb{A}\right)\oplus\mathfrak{der}\left(\mathfrak{J}_{3}\left(\mathbb{B}\right)\right)\oplus\left(\mathbb{A}'\otimes\mathfrak{J}_{3}'\left(\mathbb{B}\right)\right),
\end{equation}
where the meaning of the term $\mathbb{A}'\otimes\mathfrak{J}_{3}'\left(\mathbb{B}\right)$
will be explained later. The formula, which is not symmetric in the
entries, requires the alternativity of $\mathbb{A}$ and $\mathbb{B}$,
together with the definition of a Jordan algebra $\mathfrak{J}_{3}\left(\mathbb{B}\right)$
over the second term of the formula, i.e., the algebra $\mathbb{B}$.
With $\mathfrak{J}_{3}\left(\mathbb{B}\right)$, we denote the Jordan
algebra obtained on three-by-three Hermitian matrices $H_{3}\left(\mathbb{B}\right)$
endowed with the \emph{Jordan product}
\begin{equation}
x\cdot y=\frac{1}{2}\left(xy+yx\right),
\end{equation}
that turns the algebra commutative but not associative. In fact, it
is easy to show that in the case of $\mathbb{B}$ being an Hurwitz
algebra, then in $\mathfrak{J}_{3}\left(\mathbb{B}\right)$ the following
properties are fulfilled,
\begin{align}
x\cdot y & =y\cdot x,\\
\left(x\cdot y\right)\cdot\left(x\cdot x\right) & =x\cdot\left(y\cdot\left(x\cdot x\right)\right),
\end{align}
for every $x,y\in\mathfrak{J}_{3}\left(\mathbb{B}\right)$.

The existence of an algebra $\mathfrak{J}_{3}\left(\mathbb{B}\right)$ depends on the properties of the algebra $\mathbb{B}$ and the involution used to define the Hermitian
matrices $H_{3}\left(\mathbb{B}\right)$. As a general definition,
an involution $x\longrightarrow\overline{x}$ over the algebra $\mathbb{A}$
is called \emph{central} if $xx^{*}$ is in the center of the algebra,
i.e., elements that commutes with all the algebra, and \emph{nuclear}
if $xx^{*}$ is in the nucleus of the algebra, i.e., elements that
associate with all the algebra. {\rr In general, non-nuclear involutions prevent the Jordan identity from
holding.} In this setup, we are ready to state
a theorem \cite[Theo C.1.3]{McCrim} that strictly defines
the necessary and sufficient conditions for the existence of $\mathfrak{J}_{3}\left(\mathbb{B}\right)$.
\begin{thm}
A three-by-three matrix algebra $H_{3}\left(\mathbb{B}\right)$ under
the Jordan product $x\cdot y=\frac{1}{2}\left(xy+yx\right)$ is a
Jordan algebra $\mathfrak{J}_{3}\left(\mathbb{B}\right)$ if and only
if $\mathbb{B}$ is alternative and matrices are Hermitian with respect
to a nuclear involution.
\end{thm}

It is worth stating here a corollary that will be useful in Section 5.
\begin{cor}
\label{cor:CxO Over-the-algebra}Over the algebra of biquaternions
$\mathbb{C}\otimes\mathbb{H}$ {\rr there} exists the Jordan algebra $\mathfrak{J}_{3}\left(\mathbb{C}\otimes\mathbb{H}\right)$
with biquaternionic coefficients on the off-diagonal and real coefficients
in the diagonal, while in the case of bioctonions $\mathbb{C}\otimes\mathbb{O}$
the involution that gives Hermitian matrices $H_{3}\left(\mathbb{C}\otimes\mathbb{O}\right)$
with real elements on the diagonal is not nuclear and therefore a
$\mathfrak{J}_{3}\left(\mathbb{C}\otimes\mathbb{O}\right)$
does not exist.
\end{cor}

\begin{rem}
We used the notation $\mathfrak{J}_{3}\left(\mathbb{B}\right)$ to
identify a Jordan algebra arising from three-by-three Hermitian matrices
on the algebra $\mathbb{B}$ with respect to an involution that yields
to real elements on the diagonal. Therefore, the Corollary \emph{does
not state} that there are not possible involutions on bioctonions $\mathbb{C}\otimes\mathbb{O}$
that give rise to Jordan algebras on $H_{3}\left(\mathbb{C}\otimes\mathbb{O}\right)$.
Indeed, the involution $\widetilde{\gamma}:x\longrightarrow x^{*}=z_{0}-\sum z_{k}\text{i}_{k}^{\mathbb{O}}$
provides a Jordan algebra over Hermitian matrices that have complex
elements on the diagonal. In fact, this is the well-known Jordan algebra
$\mathfrak{J}_{3}^{\mathbb{C}}\left(\mathbb{O}\right)$, which
is the complexification of the exceptional Jordan algebra $\mathfrak{J}_{3}\left(\mathbb{O}\right)$.
\end{rem}

\begin{proof}
On the algebra of biquaternions $\mathbb{C}\otimes\mathbb{H}$ we
can define involution $\gamma$ that acts on the element
$x=z_{0}+z_{1}\text{i}_{1}^{\mathbb{H}}+z_{2}\text{i}_{2}^{\mathbb{H}}+z_{3}\text{i}_{3}^{\mathbb{H}}$
as
\begin{align}
\gamma:x &
\longrightarrow\widetilde{x}=\overline{z}_{0}-z_{1}\text{i}_{1}^{\mathbb{H}}-z_{2}\text{i}_{2}^{\mathbb{H}}-z_{3}\text{i}_{3}^{\mathbb{H}}.
\end{align}
Since the whole algebra $\mathbb{C}\otimes\mathbb{H}$ is associative, the involution is nuclear. To better understand the resulting
Jordan algebra, consider the elements on the diagonal of Hermitian
matrices. The symmetric part of the involution gives those elements,
i.e., $\left(x+\overline{x}\right)/2$, which in our case gives elements
of the form
\begin{align}
\frac{1}{2}\left(x+\widetilde{x}\right) & =\mathfrak{R}\left(z_{0}\right),
\end{align}
and therefore $\gamma$ yields a Jordan algebra $\mathfrak{J}_{3}\left(\mathbb{C}\otimes\mathbb{H}\right)$
with biquaternionic coefficients on the off-diagonal and real coefficients
on the diagonal. For the second part of the corollary, let us consider
the complex decomposition of a bioctonion
\begin{equation}
x=z_{0}+\stackrel[k=1]{7}{\sum}z_{k}\text{i}_{k}^{\mathbb{O}},
\end{equation}
and let us define the involution that leads to real elements on the
diagonal of a Hermitian matrix, i.e.,
\begin{equation}
\gamma:x\longrightarrow\widetilde{x}=\overline{z}_{0}-\stackrel[k=1]{7}{\sum}z_{k}\text{i}_{k}^{\mathbb{O}}.
\end{equation}
Then, consider the product
\begin{equation}
x\widetilde{x}=\left\Vert z_{0}\right\Vert +\left(\overline{z}_{0}-z_{0}\right)\alpha+\alpha^{2},
\end{equation}
where $\alpha=x-z_{0}\in\mathbb{O}$. It is easy to see the elements
$x\widetilde{x}$ are octonions that do not belong to the nucleus
of $\mathbb{O}$; therefore, the involution is not nuclear.
\end{proof}
\begin{cor}
In the case of the Dixon algebra $\mathbb{C}\otimes\mathbb{H}\otimes\mathbb{O}$,
the involution that gives Hermitian matrices $H_{3}\left(\mathbb{C}\otimes\mathbb{H}\otimes\mathbb{O}\right)$
with real elements on the diagonal is not nuclear and therefore $\mathfrak{J}_{3}\left(\mathbb{C}\otimes\mathbb{H}\otimes\mathbb{O}\right)$
does not exist.
\end{cor}

\section{\noun{the dixon algebra }}

The \emph{Dixon Algebra} $\mathbb{T}$ is defined as the $\mathbb{R}$-linear
tensor product of the four normed division algebras, i.e., $\mathbb{R}\otimes\mathbb{C}\otimes\mathbb{H}\otimes\mathbb{O}$
or equivalently $\mathbb{C}\otimes\mathbb{H}\otimes\mathbb{O}$ with
linear product defined by
\begin{equation}
\left(z\otimes q\otimes w\right)\left(z'\otimes q'\otimes w'\right)=zz'\otimes qq'\otimes ww',
\end{equation}
with $z,z'\in\mathbb{C}$, $q,q'\in\mathbb{H}$ and $w,w'\in\mathbb{O}$. From
the previous formula, it is evident that $\mathbb{T}$ is unital with
unit element $\text{\textbf{1}}=1\otimes1\otimes1.$

It is often easier to work with the $\mathbb{R}^{64}$ decomposition
of the algebra, where every element $t$ is of the form
\begin{align}
t & =\underset{\stackrel[\gamma=0,..,7]{\alpha=0,1}{{\scriptscriptstyle
\beta=0,..,3}}}{\sum}t^{\alpha,\beta,\gamma}\,\,\,\text{i}_{\alpha}^{\mathbb{C}}\,\text{i}_{\beta}^{\mathbb{H}}\,\,\text{i}_{\gamma}^{\mathbb{O}},\label{eq:decomposizione
64}
\end{align}
 where $t^{\alpha,\beta,\gamma}\in\mathbb{R}$, $\text{i}_{\alpha}^{\mathbb{C}}$,
$\text{i}_{\beta}^{\mathbb{H}}$ and $\text{i}_{\gamma}^{\mathbb{O}}$
are a basis for $\mathbb{C}$, $\mathbb{H}$, $\mathbb{O}$ with
\begin{align}
\text{i}_{0}^{\mathbb{C}} & =\text{i}_{0}^{\mathbb{H}}=\text{i}_{0}^{\mathbb{O}}=1,\\
\left(\text{i}_{\alpha}^{\mathbb{C}}\,\right)^{2} &
=\left(\text{i}_{\alpha}^{\mathbb{H}}\,\right)^{2}=\left(\text{i}_{\alpha}^{\mathbb{O}}\,\right)^{2}=-1,\\
\left[\text{i}_{\alpha}^{\mathbb{C}}\,,\text{i}_{\beta}^{\mathbb{H}}\,\right] &
=\left[\text{i}_{\beta}^{\mathbb{H}}\,,\text{i}_{\gamma}^{\mathbb{O}}\,\right]=\left[\text{i}_{\alpha}^{\mathbb{C}}\,,\text{i}_{\gamma}^{\mathbb{O}}\,\right]=0,\\
\left\{ \text{i}_{\alpha}^{\mathbb{H}}\,,\text{i}_{\beta}^{\mathbb{H}}\right\} \, & =\left\{
\text{i}_{\gamma}^{\mathbb{O}}\,,\text{i}_{\delta}^{\mathbb{O}}\,\right\} =0,
\end{align}
and other rules of multiplication are given in Fig. \ref{fig:Multiplication-rule-of},
when $\alpha,\beta,\gamma,\delta\neq0$. It is straightforward to
see that every element in
\begin{equation}
D=\left\{
\left(\text{i}_{\alpha}^{\mathbb{C}}\,\text{i}_{\beta}^{\mathbb{H}}\pm1\right),\,\,\left(\text{i}_{\alpha}^{\mathbb{H}}\text{i}_{\beta}^{\mathbb{O}}\pm1\right),\,\,\,\left(\text{i}_{\alpha}^{\mathbb{C}}\text{i}_{\beta}^{\mathbb{O}}\pm1\right);\alpha,\beta\neq0\right\}
\end{equation}
is a \emph{zero divisor} and therefore $\mathbb{T}$ is not a division
algebra. Moreover, the Dixon algebra is a non-associative, non-alternative,
non-flexible, and non-power-associative algebra.

\begin{figure}
\begin{centering}
\includegraphics [scale=0.06] {DixonMultiplicationRules}
\par\end{centering}
\caption{\label{fig:Multiplication-rule-of}Multiplication rule of Octonions
$\mathbb{O}$ (\emph{right}), Quaternions $\mathbb{H}$ (\emph{middle})
and Complex $\mathbb{C}$ (\emph{left}).}
\end{figure}

Nevertheless, $\mathbb{T}$ can be equipped with a \emph{norm} $N$
from $\mathbb{T}$ in $\mathbb{R}$ given by
\begin{equation}
N\left(t\right)=\underset{\stackrel[\gamma=0,..,7]{\alpha=0,1}{{\scriptscriptstyle
\beta=0,..,3}}}{\sum}\left(t^{\alpha,\beta,\gamma}\right)^{2},\label{eq:Norma di T}
\end{equation}
where is $t$ an element of the algebra with decomposition given in
(\ref{eq:decomposizione 64}). Along with the norm we have its associated
\emph{polar form} $\left\langle \cdot,\cdot\right\rangle $ given
by the symmetric bilinear form
\begin{equation}
2\left\langle t_{1},t_{2}\right\rangle =N\left(t_{1}+t_{2}\right)-N\left(t_{1}\right)-N\left(t_{2}\right).
\end{equation}
Finally, one might consider the conjugation on $\mathbb{T}$ given
by $\overline{t}=2\left\langle t,\text{\textbf{1}}\right\rangle \text{\textbf{1}}-t$, even
though it will not be necessary for our purposes.

\section{\noun{three lie algebras}}

In this section, we define three Lie algebras, i.e., $\mathfrak{A}_{1}$,
$\mathfrak{A}_{2}$ and $\mathfrak{A}_{3}$, that correspond to the isometry
groups needed for the definition of the three Dixon-Rosenfeld planes.
The process used to construct these algebras is well known and is
linked to Tits-Freudenthal's magic square \cite{Freud 1965,Tits},
which is an array of Lie algebras usually obtained from two Hurwitz
algebras, but that can be extended to composition algebras and even
to alternative algebras \cite{BS,Squaring-Magic}.

\paragraph*{A few remarks on magic squares} \,

An entry of the Tits-Freudenthal's magic square $\mathcal{L}_{3}\left(\mathbb{A},\mathbb{B}\right)$
is a Lie algebra that can be realized from three different
recipes based on two algebras $\mathbb{A}$ and $\mathbb{B}$.
We will briefly present them here without the explicit definition
of the Lie brackets found in \cite{BS}. These three magic
square formulas are not the only known formulas of this kind. It is
worth mentioning the general formula of Santander-Herranz based on
Hurwitz algebras \cite{Santander}, the Atsuyama construction \cite{Atsuyama1}
for any composition algebra, and the Elduque magic square designed
for symmetric composition algebras, i.e., composition algebras that
are flexible instead of unital and alternative \cite{EldMS1,EldMS2}.

A general requirement for all the formulas, except Tits' formula,
is that the two algebras $\mathbb{A}$ and $\mathbb{B}$ are composition
algebras. In this case, a symmetric formula on the two algebras, i.e.,
$\mathcal{L}_{3}\left(\mathbb{A},\mathbb{B}\right)=\mathcal{L}_{3}\left(\mathbb{B},\mathbb{A}\right)$,
was obtained by Vinberg \cite{Vinberg}
\begin{equation}
\mathcal{L}_{3}\left(\mathbb{A},\mathbb{B}\right)=\mathfrak{der}\left(\mathbb{A}\right)\oplus\mathfrak{der}\left(\mathbb{B}\right)\oplus\mathfrak{sa}_{3}\left(\mathbb{A}\otimes\mathbb{B}\right),
\end{equation}
involving only the derivation algebras of $\mathbb{A}$ and $\mathbb{B}$
together with the three-by-three antisymmetric matrices over $\mathbb{A}\otimes\mathbb{B}$.
More recently, Barton and Sudbury \cite{BS} utilized the
triality algebras $\mathfrak{tri}\left(\mathbb{A}\right)$ and $\mathfrak{tri}\left(\mathbb{B}\right)$
together with three copies of the tensor product $\mathbb{A}\otimes\mathbb{B}$,
i.e.,
\begin{equation}
\mathcal{L}_{3}\left(\mathbb{A},\mathbb{B}\right)=\mathfrak{tri}\left(\mathbb{A}\right)\oplus\mathfrak{tri}\left(\mathbb{B}\right)\oplus3\left(\mathbb{A}\otimes\mathbb{B}\right).
\end{equation}

While the previous formulas require $\mathbb{A}$
and $\mathbb{B}$ to be composition algebras, Tits' formula \cite{Tits} is more general because it only requires $\mathbb{A}$ to be
alternative and $\mathfrak{J}_{n}\left(\mathbb{B}\right)$ to be a Jordan algebra,
\begin{equation}
\mathcal{L}_{n}\left(\mathbb{A},\mathbb{B}\right)=\mathfrak{der}\left(\mathbb{A}\right)\oplus\mathfrak{der}\left(\mathfrak{J}_{n}\left(\mathbb{B}\right)\right)\oplus\left(\mathbb{A}'\otimes\mathfrak{J}_{n}'\left(\mathbb{B}\right)\right),
\end{equation}
where the meaning of the tensor product $\mathbb{A}'\otimes\mathfrak{J}_{n}'\left(\mathbb{B}\right)$
will be explained soon.

Tits' construction is not manifestly symmetric. Since it was the
first formula discovered, its symmetry in the case of $n=2,3$ when $\mathbb{A}$
and $\mathbb{B}$ are Hurwitz algebras appeared to be ``magic'', resulting in the so-called ``magic square’’. Nevertheless,
 the symmetry is not always ensured, since $\mathbb{A}=\mathbb{O}$ and $\mathbb{B}\in\left\{ \mathbb{R},\mathbb{C},\mathbb{H}\right\} $
with $n>3$ give
\begin{equation}
\mathcal{L}_{n}\left(\mathbb{O},\mathbb{B}\right)=\mathfrak{der}\left(\mathbb{O}\right)\oplus\mathfrak{der}\left(\mathfrak{J}_{n}\left(\mathbb{B}\right)\right)\oplus\left(\mathbb{O}'\otimes\mathfrak{J}_{n}'\left(\mathbb{B}\right)\right).
\end{equation}
This is a perfectly reasonable construction, since both $\mathbb{O}$
and $\mathbb{B}$ are alternative algebras and the Jordan algebra
$\mathfrak{J}_{n}\left(\mathbb{B}\right)$ is well defined for every
$n>3$. Alternatively, there are no Jordan algebras $\mathfrak{J}_{n}\left(\mathbb{O}\right)$
over the octonions with $n>3$. In this case, $\mathcal{L}_{n}\left(\mathbb{B},\mathbb{O}\right)$
cannot even be defined.

\paragraph*{Three Lie algebras for the isometries of Dixon-Rosenfeld planes} \,

Given the Dixon algebra, we are interested in extending Rosenfeld's
approach \cite{Rosenf98,Rosenfeld Group2-1}, where the Lie algebra
from Tits' formula was considered as
the Lie algebra of the isometry group of a projective plane over the
tensor product of the two algebras. This characterization of the projective
plane $\left(\mathbb{A}\otimes\mathbb{B}\right)P^{2}$ as a symmetric
space from an isometry group whose algebra was given by the entry
of the magic square $\mathcal{L}_{3}\left(\mathbb{A},\mathbb{B}\right)$,
i.e.,
\begin{equation}
\mathcal{L}_{3}\left(\mathbb{A},\mathbb{B}\right)\cong\mathfrak{isom}\left(\left(\mathbb{A}\otimes\mathbb{B}\right)P^{2}\right),
\end{equation}
was already well known by Cartan in the non-octonionic
case \cite{Cartan}. Rosenfeld postulated the existence of a projective plane $\left(\mathbb{A}\otimes\mathbb{B}\right)P^{2}$
for the octonionic case, which was proved to be incorrect in a strict
projective sense, but true in a ``wider sense'' as rigorously investigated
by Atsuyama \cite{Atsuyama,Atiyah Bernd}. In this wider sense, the
approach was reversed. Instead of defining a projective plane
and then studying its isometry group, the projective plane was directly
defined as a homogeneous space starting from a specific isometry group
and only then its projective properties were studied \cite{Corr Rosen,RealF}.

This is the line of thought we will adopt in defining
three Dixon-Rosenfeld planes. Therefore, we will write the Dixon algebra
$\mathbb{T}=\mathbb{C}\otimes\mathbb{H}\otimes\mathbb{O}$ as a tensor
product of the form $\mathbb{A}\otimes\mathbb{B}$ to obtain
the Lie algebra $\mathcal{L}_{3}\left(\mathbb{A},\mathbb{B}\right)$
of the isometry group. Since the only unital composition algebras
are $\mathbb{R},\mathbb{C},\mathbb{H},\mathbb{O},\mathbb{C}_{s},\mathbb{H}_{s},\mathbb{O}_{s}$
(cf.~\cite{ElDuque Comp}) and since all other formulas make essential
use of the the algebra being composition, this leaves Tits' formula
as the only formula suitable for our purposes, since it's the only
one that only requires alternativity on both $\mathbb{A}$ and $\mathbb{B}$.
Moreover, since $\mathbb{H}\otimes\mathbb{O}$ is not alternative,
the possible candidates for $\mathfrak{isom}\left(\mathbb{T}P^{2}\right)$
can \emph{a priori} only be related with the following four different
$\mathbb{A}$ and $\mathbb{B}$, i.e.,
\begin{equation}
\begin{array}{lccl}
I: & \left(\mathbb{A},\mathbb{B}\right) & = & \left(\mathbb{C}\otimes\mathbb{H},\mathbb{O}\right),\\
II: & \left(\mathbb{A},\mathbb{B}\right) & = & \left(\mathbb{O},\mathbb{C}\otimes\mathbb{H}\right),\\
III: & \left(\mathbb{A},\mathbb{B}\right) & = & \left(\mathbb{C}\otimes\mathbb{O},\mathbb{H}\right).
\end{array}\label{3-cases}
\end{equation}
A fourth case $\left(\mathbb{A},\mathbb{B}\right)=\left(\mathbb{H},\mathbb{C}\otimes\mathbb{O}\right)$ requires the existence of a Jordan algebra $\mathfrak{J}_{3}\left(\mathbb{C}\otimes\mathbb{O}\right)$, which is not possible (see Corollary \ref{cor:CxO Over-the-algebra}).\footnote{We intentionally exclude the cases where the involution on Hermitian
matrices doesn't involve the whole algebra, i.e., the case that would
involve $\mathfrak{J}_{3}^{\mathbb{C}}\left(\mathbb{O}\right)$ and
$\mathfrak{J}_{3}^{\mathbb{C}}\left(\mathbb{H}\right)$ for which
the analogy with Rosenfeld construction does not hold.} Therefore, we are left with only three different possibilities,
\begin{align}
\mathfrak{A}_{1}=\mathcal{L}_{3}\left(\mathbb{C}\otimes\mathbb{H},\mathbb{O}\right) & ,\\
\mathfrak{A}_{2}=\mathcal{L}_{3}\left(\mathbb{O},\mathbb{C}\otimes\mathbb{H}\right) & ,\\
\mathfrak{A}_{3}=\mathcal{L}_{3}\left(\mathbb{C}\otimes\mathbb{O},\mathbb{H}\right) & .
\end{align}

To define these algebras, we follow \cite[Sec.~3]{BS}
in notation. Given an algebra $\mathbb{A}$, we define
\begin{equation}
X'=X-\frac{1}{n}\text{Tr}\left(X\right)\text{\textbf{1}},\label{eq:apostrofo1}
\end{equation}
as the projection of an element of the algebra in the subspace orthogonal
to the identity denoted as $\text{\textbf{1}}$. We then define $\mathfrak{J}'_{3}\left(\mathbb{B}\right)$ as
the algebra obtained by such elements with the product $\bullet$
given by the projection back on the subspace orthogonal to the identity
of the Jordan product, i.e.,
\begin{equation}
X'\bullet Y'=X'\cdot Y'-\frac{4}{3}\left\langle X',Y'\right\rangle \text{\textbf{1}},\label{eq:apostrofo2}
\end{equation}
where, as usual we intended $X\cdot Y=XY+YX$ and $\left\langle X,Y\right\rangle =\nicefrac{1}{2}\text{Tr}\left(X\cdot Y\right)$
for every $X,Y\in\mathfrak{J}_{3}\left(\mathbb{B}\right)$. With the
previous notation, Tits' formula defines a Lie algebra structure on
the vector space
\begin{equation}
\mathcal{L}_{3}\left(\mathbb{A},\mathbb{B}\right)=\mathfrak{der}\left(\mathbb{A}\right)\oplus\mathfrak{der}\left(\mathfrak{J}_{3}\left(\mathbb{B}\right)\right)\oplus\left(\mathbb{A}'\otimes\mathfrak{J}_{3}'\left(\mathbb{B}\right)\right),
\end{equation}
endowed with the following brackets:
\begin{enumerate}
\item The usual brackets on the Lie subalgebra
    $\mathfrak{der}\left(\mathbb{A}\right)\oplus\mathfrak{der}\left(\mathfrak{J}_{3}\left(\mathbb{B}\right)\right)$.
\item When $a\in\mathfrak{der}\left(\mathbb{A}\right)\oplus\mathfrak{der}\left(\mathfrak{J}_{3}\left(\mathbb{B}\right)\right)$
and $A\in\mathbb{A}'\otimes\mathfrak{J}_{3}'\left(\mathbb{B}\right)$
then
\begin{equation}
\left[a,A\right]=a\left(A\right).
\end{equation}
\item When $a\otimes A,b\otimes B\in\mathbb{A}'\otimes\mathfrak{J}_{3}'\left(\mathbb{B}\right)$
then
\begin{equation}
\left[a\otimes A,b\otimes B\right]=\frac{1}{3}\left\langle A,B\right\rangle D_{a,b}-\left\langle a,b\right\rangle
\left[L_{A},L_{B}\right]+\frac{1}{2}\left[a,b\right]\otimes\left(A\bullet B\right),
\end{equation}
 where $L_{x}$ and $R_{x}$ are the left and right actions on the
algebra and $D_{x,y}$ is given by
\begin{equation}
D_{x,y}=\left[L_{x},L_{y}\right]+\left[L_{x},R_{y}\right]+\left[R_{x},R_{y}\right].
\end{equation}
\end{enumerate}
Applying the previous formula to the allowed cases in (\ref{3-cases}) gives the following:
\begin{description}
\item [{I:}] For $\mathbb{A}=\mathbb{C}\otimes\mathbb{H}$ and $\mathbb{B}=\mathbb{O}$,
the vector space is given by
\begin{equation}
\mathfrak{A}_{1}=\mathcal{L}_{3}\left(\mathbb{C}\otimes\mathbb{H},\mathbb{O}\right)=\mathfrak{su}_{2}\oplus\mathfrak{f}_{4}\oplus\left(\mathbb{\left(\mathbb{C}\otimes\mathbb{H}\right)}'\otimes\mathfrak{J}_{3}'\left(\mathbb{O}\right)\right),\label{eq:Alg1}
\end{equation}
where $\mathbb{\left(\mathbb{C}\otimes\mathbb{H}\right)}'\otimes\mathfrak{J}_{3}'\left(\mathbb{O}\right)$
corresponds to the representation $\left(\boldsymbol{7},\boldsymbol{26}\right)$
of $\mathfrak{su}_{2}\oplus\mathfrak{f}_{4}$. The Lie algebra $\mathfrak{A}_{1}$
has therefore real dimension $237=3+52+7\times26$.
\item [{II:}] For $\mathbb{A}=\mathbb{O}$ and $\mathbb{B}=\mathbb{C}\otimes\mathbb{H}$,
we have the real vector space
\begin{equation}
\mathfrak{A}_{2}=\mathcal{L}_{3}\left(\mathbb{O},\mathbb{C}\otimes\mathbb{H}\right)=\mathfrak{g}_{2}\oplus\mathfrak{su}_{2}\oplus\left(\mathbb{O}'\otimes\mathfrak{J}_{3}'\mathbb{\left(\mathbb{C}\otimes\mathbb{H}\right)}\right),\label{eq:Alg2}
\end{equation}
where $\mathbb{O}'\otimes\mathfrak{J}_{3}'\mathbb{\left(\mathbb{C}\otimes\mathbb{H}\right)}$
gives the representation $\left(7,3\cdot\boldsymbol{7}+5\cdot\boldsymbol{1}\right)$
of $\mathfrak{g}_{2}\oplus\mathfrak{su}_{2}$. The Lie algebra $\mathfrak{A}_{2}$
has real dimension $199=14+3+7\times26$.
\item [{III:}] For $\mathbb{A}=\mathbb{C}\otimes\mathbb{O}$ and
$\mathbb{B}=\mathbb{H}$, Tits' construction gives the real vector
space
\begin{equation}
\mathfrak{A}_{3}=\mathcal{L}_{3}\left(\mathbb{C}\otimes\mathbb{O},\mathbb{H}\right)=\mathfrak{g}_{2}\oplus\mathfrak{sp}_{6}\oplus\left(\mathbb{\left(\mathbb{C}\otimes\mathbb{O}\right)}'\otimes\mathfrak{J}_{3}'\left(\mathbb{H}\right)\right),\label{eq:Alg3}
\end{equation}
where $\mathbb{\left(\mathbb{C}\otimes\mathbb{O}\right)}'\otimes\mathfrak{J}_{3}'\left(\mathbb{H}\right)$
is the $\left(2\cdot\mathbf{7}+\mathbf{1},\mathbf{14}\right)$ representation
of $\mathfrak{g}_{2}\oplus\mathfrak{sp}_{6}$. The Lie algebra $\mathfrak{A}_{3}$
has dimension $245=14+21+15\times14$.
\end{description}

\section{\noun{Three Dixon-Rosenfeld planes}}

In this section, we define three \emph{coset manifolds} of real dimension
$128$ whose isometry groups are the Lie algebras
$\mathfrak{A}_{1}$, $\mathfrak{A}_{2}$, and $\mathfrak{A}_{3}$.
In the framework of coset manifolds, the Lie algebra of the isometry
group $\mathfrak{Lie}\left(G\right)$, the isometry group
itself $G$, and the isotropy group $H$ are the only objects required to specify a coset manifold and its metric properties.
The definitions of the Dixon-Rosenfeld planes given here are
self-contained and include both geometrical and metrical properties.

\paragraph*{A few remarks on coset manifolds} \,

It is well known from (pseudo-)Riemannian geometry that the condition
for a diffeomorphism $\phi$ to be an isometry of a Riemannian manifold
$\left(\mathcal{M},g\right)$ is the condition for its components
$\phi^{\mu}\left(x\right)\simeq x^{\mu}+k^{\mu}\left(x\right)$ to
solve the equation
\begin{equation}
\nabla_{\mu}k_{\nu}+\nabla_{\nu}k_{\mu}=0.
\end{equation}
Hence, the vector $k=\left(k^{1},...,k^{n}\right)$ must
be a \emph{Killing vector} for the metric $g$. It is also well-known
that Killing vector fields form a finite dimensional Lie algebra $\mathfrak{isom}\left(\mathcal{M},g\right)$
which is, in fact, the Lie algebra of the manifold's isometry group.
Reversing the argument, we are now interested in constructing (pseudo-)Riemannian manifolds that have a prescribed isometry algebra.

\emph{Coset manifolds} arise from coset spaces over a Lie group $G$
given by an equivalence relation of the type
\begin{equation}
g\sim g'\Longleftrightarrow gh=g',
\end{equation}
where $g,g'\in G$, $h\in H$, and $H$ is a subgroup of $G$. In
this case, the coset space $G/H$, obtained from the equivalence classes
$gH$, inherits a manifold structure from $G$ and is therefore a
manifold of dimension
\begin{equation}
\text{dim}\left(G/H\right)=\text{dim}\left(G\right)-\text{dim}\left(H\right).
\end{equation}
Moreover, $G/H$ can be endowed with invariant metrics such that all
elements of the original group $G$ are isometries of the constructed
metric \cite{Fre and Fedotov}. More specifically, the generators
of the infinitesimal isometries are the Killing vector fields as desired.
The structure constants of the Lie algebra of $G$,
i.e., $\mathfrak{Lie}\left(G\right)$, completely define the metric
and therefore all the metric-dependent tensors, such as the curvature
tensor, the Ricci tensor, etc. Finally, by construction, the coset
space $G/H$ is a homogeneous manifold, i.e., the group $G$ acts transitively,
and its \emph{isotropy subgroup} is precisely $H$, i.e., the group
$H$ is such that for any given point $p$ in the manifold, $hp=p$.

\paragraph*{The Dixon-Rosenfeld planes} \,

Given the Lie algebras $\mathfrak{A}_{1}$, $\mathfrak{A}_{2}$, and $\mathfrak{A}_{3}$
obtained from Tits' formula in (\ref{eq:Alg1}), (\ref{eq:Alg2})
and (\ref{eq:Alg3}) we are allowed to define three different coset
spaces having the previous as Lie algebras of the isometry group. More
specifically, we will use $\mathfrak{A}_{1}$ and $\mathfrak{A}_{3}$
as $\mathfrak{isom}\left(\mathbb{T}P_{I}^{2}\right)$ and $\mathfrak{isom}\left(\mathbb{T}P_{III}^{2}\right)$
respectively, while the algebra $\mathfrak{A}_{2}$ is not suitable
to be used directly as $\mathfrak{isom}\left(\mathbb{T}P_{II}^{2}\right)$
and will need an enhancement $\mathfrak{A}_{2}^{\text{enh}}$. Having
defined the isometry algebras, knowing that every Dixon-Rosenfeld
plane is a 128-dimensional plane, i.e., dim$_{\mathbb{R}}\mathbb{T}P^{2}=2$dim$_{\mathbb{R}}\mathbb{T}=128$,
we will find three suitable isotropy algebras and, therefore, three
suitable isotropy groups to define the three coset manifolds
of the desired dimension. The three Dixon-Rosenfeld planes will then
be realized as homogeneous spaces, whose coset algebra $\mathfrak{c}\left(\mathbb{T}P^{2}\right)$
is the complementary of the Lie algebra $\mathfrak{isot}\left(\mathbb{T}P^{2}\right)$
within the larger Lie algebra $\mathfrak{isom}\left(\mathbb{T}P^{2}\right)$
in the Cartan decomposition
\begin{equation}
\mathfrak{isom}\left(\mathbb{T}P^{2}\right)=\mathfrak{isot}\left(\mathbb{T}P^{2}\right)\oplus\mathfrak{c}\left(\mathbb{T}P^{2}\right).\label{eq:isom=00003Disot+c}
\end{equation}
The Lie algebra of the coset space, namely $\mathfrak{c}\left(\mathbb{T}P^{2}\right)$,
carries a generally reducible module of $\mathfrak{isot}\left(\mathbb{T}P^{2}\right)$,
which provides a projective representation of $\mathbb{T\oplus T\simeq T}^{\oplus2}$
itself, and determines the tangent space $T\left(\mathbb{T}P^{2}\right)$
of the plane $\mathbb{T}P^{2}$ under consideration (see below). Finally,
it is worth noting that in general, the follow holds
\begin{equation}
\left[\mathfrak{c}\left(\mathbb{T}P^{2}\right),\mathfrak{c}\left(\mathbb{T}P^{2}\right)\right]\subset\mathfrak{isom}\left(\mathbb{T}P^{2}\right),
\end{equation}
while the condition
$\left[\mathfrak{c}\left(\mathbb{T}P^{2}\right),\mathfrak{c}\left(\mathbb{T}P^{2}\right)\right]\subset\mathfrak{isot}\left(\mathbb{T}P^{2}\right)$
is satisfied only when $\mathbb{T}P^{2}$ is a \textit{symmetric}
coset, which we will show is not the case.

\paragraph*{The first Dixon-Rosenfeld plane} \,

We now define the first Dixon-Rosenfeld plane $\mathbb{T}P_{I}^{2}$,
starting from the Lie algebra of its isometry group $\mathfrak{A}_{1}$, which in this case is $\mathcal{L}_{3}\left(\mathbb{C}\otimes\mathbb{H},\mathbb{O}\right)$.
Applying Tits' formula, we obtain
\begin{align}
\mathfrak{isom}\left(\mathbb{T}P_{I}^{2}\right): & =\mathfrak{A}_{1}\\
\notag & =\mathfrak{su}_{2}\oplus\mathfrak{f}_{4}\oplus\left(\mathbf{7},\mathbf{26}\right),
\end{align}
{\rr thus} giving the projective representation of $\mathbb{T}^{\oplus2}$
given by
\begin{equation}
T\left(\mathbb{T}P_{I}^{2}\right)\simeq2\cdot\left(\mathbf{7}+\mathbf{1},\mathbf{7}+\mathbf{1}\right)~\text{of~}\mathfrak{su}_{2}\oplus\mathfrak{g}_{2},\label{TI-plane}
\end{equation}
where $\mathfrak{su}_{2}\oplus\mathfrak{g}_{2}$ is found in
Tits' formula from $\mathfrak{der}\left(\mathbb{C}\otimes\mathbb{H}\right)\oplus\mathfrak{der}\left(\mathbb{O}\right)$.
By iterated branchings of $\mathfrak{isom}\left(\mathbb{T}P_{I}^{1}\right)$,
one obtains
\begin{eqnarray*}
\mathfrak{su}_{2}\oplus\mathfrak{f}_{4}\oplus\left(\mathbf{7},\mathbf{26}\right) & = &
\mathfrak{su}_{2}\oplus\mathfrak{so}_{9}\oplus\left(\mathbf{7},\mathbf{1+9+16}\right)\oplus\left(\mathbf{1},\mathbf{16}\right)\\
 & = &
 \mathfrak{su}_{2}\oplus\mathfrak{so}_{8}\oplus\left(\mathbf{7},\mathbf{1+1+8}_{v}\mathbf{+8}_{s}+\mathbf{8}_{c}\right)\oplus\left(\mathbf{1},\mathbf{8}_{v}\mathbf{+8}_{s}+\mathbf{8}_{c}\right)\\
 & = &
 \mathfrak{su}_{2}\oplus\mathfrak{so}_{7}\oplus\left(\mathbf{7},3\cdot\mathbf{1+7+}2\cdot\mathbf{8}\right)\oplus\left(\mathbf{1},2\cdot\mathbf{8}+2\cdot\mathbf{7+1}\right)\\
 & = &
 \mathfrak{su}_{2}\oplus\mathfrak{g}_{2}\oplus\left(\mathbf{7},5\cdot\mathbf{1+}3\cdot\mathbf{7}\right)\oplus\left(\mathbf{1},5\cdot\mathbf{7+}3\cdot\mathbf{1}\right),
\end{eqnarray*}
and, since $\mathfrak{isom}\left(\mathbb{T}P_{I}^{2}\right)\simeq\mathfrak{isot}\left(\mathbb{T}P_{I}^{2}\right)\oplus
T\left(\mathbb{T}P_{I}^{2}\right)$,
the following isotropy algebra is found,
\begin{equation}
\mathfrak{isot}\left(\mathbb{T}P_{I}^{2}\right)\simeq\mathfrak{su}_{2}\oplus\mathfrak{g}_{2}\oplus\left(\mathbf{7},3\cdot\mathbf{1+7}\right)\oplus\left(\mathbf{1},3\cdot\mathbf{7+1}\right).\label{isot1-plane}
\end{equation}
From the previous considerations, we obtain a realization
of the first Dixon projective plane $\mathbb{T}P_{I}^{2}$ as the
homogeneous space given by
\begin{equation}
\mathbb{T}P_{I}^{2}\coloneqq\frac{SU_{2}\times F_{4}\ltimes\left(\mathbf{7},\mathbf{26}\right)}{SU_{2}\times
G_{2}\ltimes\left(\left(\mathbf{7},3\cdot\mathbf{1+7}\right)\times\left(\mathbf{1},3\cdot\mathbf{7+1}\right)\right)}.\label{DR1-plane}
\end{equation}
A simple dimensional check shows that the plane has the desired
dimension, twice that of the Dixon algebra, i.e.,
\begin{align}
\text{dim}\left(\mathbb{T}P_{I}^{2}\right) & =3+52+182-3-14-70-22\\
\notag & =128=2\text{dim}\mathbb{T}.
\end{align}
Finally, it is worth noting that the coset (\ref{DR1-plane}) is non-symmetric since
\begin{equation}
\left[\mathfrak{c}\left(\mathbb{T}P_{I}^{2}\right),\mathfrak{c}\left(\mathbb{T}P_{I}^{2}\right)\right]\simeq4\cdot\left(\mathbf{7}+\mathbf{1},\mathbf{7}+\mathbf{1}\right)\otimes_{a}\left(\mathbf{7}+\mathbf{1},\mathbf{7}+\mathbf{1}\right)\nsubseteq\mathfrak{isot}\left(\mathbb{T}P_{I}^{2}\right).
\end{equation}

\paragraph*{The second Dixon-Rosenfeld plane} \,

We now proceed with identifying the isometry algebra of the second Dixon-Rosenfeld
plane, i.e., $\mathfrak{isom}\left(\mathbb{T}P_{II}^{2}\right)$. Our
starting point is given by $\mathfrak{A}_{2}=\mathcal{L}_{3}\left(\mathbb{O},\mathbb{C}\otimes \mathbb{H}\right)$.
In this case, the projective representation of $\mathbb{T}^{\oplus2}$
is still given by (\ref{TI-plane}), i.e.,
\begin{equation}
T\left(\mathbb{T}P_{II}^{2}\right)\simeq2\cdot\left(\mathbf{7}+\mathbf{1},\mathbf{7}+\mathbf{1}\right)~\text{of~}\mathfrak{g}_{2}\oplus\mathfrak{su}_{2},\label{TII-plane}
\end{equation}
and the use of Tits' formula gives
\begin{equation}
\mathfrak{A}_{2} = \mathfrak{g}_{2}\oplus\mathfrak{su}_{2}\oplus\left(\mathbf{7},3\cdot\mathbf{7+}5\cdot\mathbf{1}\right).
\end{equation}
Unfortunately, considering (\ref{TI-plane}) and (\ref{TII-plane}),
we notice that
\begin{equation}
\mathbb{T}^{\oplus2}\nsubseteq\mathfrak{g}_{2}\oplus\mathfrak{su}_{2}\oplus\left(\mathbf{7},3\cdot\mathbf{7+}5\cdot\mathbf{1}\right),\label{not2}
\end{equation}
because the $\mathfrak{g}_{2}$-singlets of (\ref{TI-plane}) are
missing in the r.h.s. of (\ref{not2}). Thus, if we want to define
a Dixon-Rosenfeld plane starting from the Lie algebra $\mathfrak{isom}\left(\mathbb{T}P_{II}^{1}\right)$
determined by Tits' formula, we must necessarily add \textit{at least}
by a set of generators sitting in the $2\cdot\mathbf{7+}2\cdot\mathbf{1}$
of $\mathfrak{su}_{2}$, such that the ``minimal''\ enhancement
of $\mathfrak{A}_{2}$, i.e., $\mathfrak{A}_{2}^{\text{enh}}$, is
suitable to be candidate of the isometry algebra of a Dixon-Rosenfeld
plane $\mathfrak{isom}\left(\mathbb{T}P_{II}^{2}\right)$. We thus
have
\begin{eqnarray}
\mathfrak{isom}\left(\mathbb{T}P_{II}^{2}\right) & \coloneqq & \mathfrak{A}_{2}^{\text{enh}}\nonumber \\
 & = &
 \mathfrak{g}_{2}\oplus\mathfrak{su}_{2}\oplus\left(\mathbf{7},3\cdot\mathbf{7+}5\cdot\mathbf{1}\right)\oplus2\cdot\left(\mathbf{1},\mathbf{7+1}\right)\\
 & \simeq &
 \mathfrak{g}_{2}\oplus\mathfrak{su}_{2}\oplus\left(\mathbf{7},\mathbf{7+}3\cdot\mathbf{1}\right)\oplus\mathbb{T}^{\oplus2}.\label{en1-plane}
\end{eqnarray}
Consequently, from (\ref{eq:isom=00003Disot+c}), (\ref{TII-plane})
and the ``minimally'' enhanced Lie algebra $\mathfrak{A}_{2}^{\text{enh}}$
in (\ref{en1-plane}), we have the isotropy algebra and the coset
algebra given by
\begin{eqnarray}
\mathfrak{isot}\left(\mathbb{T}P_{II}^{2}\right) & \simeq &
\mathfrak{g}_{2}\oplus\mathfrak{su}_{2}\oplus\left(\mathbf{7},\mathbf{7+}3\cdot\mathbf{1}\right),\label{isot2-plane}\\
\mathfrak{c}\left(\mathbb{T}P_{II}^{2}\right) & \simeq & T\left(\mathbb{T}P_{II}^{2}\right).
\end{eqnarray}
Therefore, one obtains the following characterization of the Dixon
projective plane $\mathbb{T}P_{II}^{2}$ as homogeneous space
\begin{equation}
\mathbb{T}P_{II}^{2}\simeq\frac{G_{2}\times
SU_{2}\ltimes\left(\left(\mathbf{7},3\cdot\mathbf{7+}5\cdot\mathbf{1}\right)\times2\cdot\left(\mathbf{1},\mathbf{7+1}\right)\right)}{G_{2}\times
SU_{2}\ltimes\left(\mathbf{7},\mathbf{7+}3\cdot\mathbf{1}\right)}.\label{DR2-plane}
\end{equation}
The dimensional count gives back the desired dimension
\begin{align}
\text{dim}\mathbb{T}P_{II}^{2} & =14+3+7\cdot26+16-14-3-70\\
\notag & =128=2\text{dim}\mathbb{T},
\end{align}
and, once again, the coset (\ref{DR2-plane}) is not symmetric, since
\begin{equation}
\left[\mathfrak{c}\left(\mathbb{T}P_{II}^{1}\right),\mathfrak{c}\left(\mathbb{T}P_{II}^{1}\right)\right]\simeq\left(\mathbf{7}+\mathbf{1},\mathbf{7}+\mathbf{1}\right)\otimes_{a}\left(\mathbf{7}+\mathbf{1},\mathbf{7}+\mathbf{1}\right)\nsubseteq\mathfrak{isot}\left(\mathbb{T}P_{II}^{1}\right).
\end{equation}

\paragraph*{The third Dixon-Rosenfeld plane} \,

Finally, we proceed with defining the isometry algebra of the third Dixon-Rosenfeld plane, i.e.,
$\mathfrak{iso}\left(\mathbb{T}P_{III}^{2}\right)\cong\mathfrak{A}_{3}=\mathcal{L}_{3}\left(\mathbb{C}\otimes\mathbb{O},\mathbb{H}\right).$
Here, the Dixon algebra $\mathbb{T}$ is again a $\mathfrak{g}_{2}\oplus\mathfrak{su}_{2}$-module,
but a different one with respect to the cases I and II, namely
\begin{equation}
\mathbb{T}^{\oplus2}\simeq
T\left(\mathbb{T}P_{III}^{2}\right)\simeq4\cdot\left(\mathbf{7}+\mathbf{1},\mathbf{3}+\mathbf{1}\right)~\text{of~}\mathfrak{g}_{2}\oplus\mathfrak{su}_{2},\label{TIII-plane}
\end{equation}
where $\mathfrak{g}_{2}\oplus\mathfrak{su}_{2}$ is found
from $\mathfrak{der}\left(\mathbb{C}\otimes\mathbb{O}\right)\oplus\mathfrak{der}\left(\mathbb{H}\right)$.
In this case, the isometry algebra is given directly by the $\mathfrak{A}_{3}$
algebra resulting from Tits' formula, i.e.,
\begin{align}
\mathfrak{isom}\left(\mathbb{T}P_{III}^{2}\right) &:=\mathfrak{A}_{3}\label{isom3-plane}\\
\notag & =\mathfrak{g}_{2}\oplus\mathfrak{sp}_{6}\oplus\left(2\cdot\mathbf{7}+\mathbf{1},\mathbf{14}\right),
\end{align}
where$\mathbf{14}\equiv\wedge_{0}^{2}\mathbf{6}\text{~of~}\mathfrak{sp}_{6}.$
Therefore, by iterated branchings of $\mathfrak{isom}\left(\mathbb{T}P_{III}^{2}\right)$,
one obtains
\begin{eqnarray*}
\mathfrak{g}_{2}\oplus\mathfrak{sp}_{6}\oplus\left(2\cdot\mathbf{7}+\mathbf{1},\mathbf{14}\right) & = &
\mathfrak{g}_{2}\oplus\mathfrak{su}_{2,I}\oplus\mathfrak{su}_{2,II}\oplus\left(\mathbf{1},\mathbf{3,5}\right)\oplus\left(2\cdot\mathbf{7}+\mathbf{1},\mathbf{1,5}\right)\oplus\left(2\cdot\mathbf{7}+\mathbf{1},\mathbf{3,3}\right)\\
 & = &
 \mathfrak{g}_{2}\oplus\mathfrak{su}_{2,I}\oplus\left(\mathbf{1},3\cdot\mathbf{1}+5\cdot\mathbf{3}\right)\oplus\left(2\cdot\mathbf{7}+\mathbf{1},5\cdot\mathbf{1}+3\cdot\mathbf{3}\right)\\
\notag & = &
\mathfrak{g}_{2}\oplus\mathfrak{su}_{2}\oplus\left(\mathbf{7+1},4\cdot\mathbf{1}+2\cdot\mathbf{3}\right)\oplus2\cdot\left(\left(\mathbf{7},\mathbf{1}\right)\oplus\left(\mathbf{1},\mathbf{3}\right)\right)\oplus\mathbb{T}^{\oplus2}
\end{eqnarray*}
where the isotropy algebra is given by
\begin{align}
\mathfrak{isot}\left(\mathbb{T}P_{III}^{2}\right)\coloneqq &
\mathfrak{g}_{2}\oplus\mathfrak{su}_{2}\oplus\left(\mathbf{7+1},4\cdot\mathbf{1}+2\cdot\mathbf{3}\right)\label{isot3-plane}\\
\notag & \oplus2\cdot\left(\left(\mathbf{7},\mathbf{1}\right)\oplus\left(\mathbf{1},\mathbf{3}\right)\right).
\end{align}
Consequently, the
Dixon projective plane $\mathbb{T}P_{III}^{2}$ is realized as the homogeneous space:
\begin{equation}
\mathbb{T}P_{III}^{2}\simeq\frac{G_{2}\times Sp_{6}\ltimes\left(2\cdot\mathbf{7}+\mathbf{1},\mathbf{14}\right)}{G_{2}\times
SU_{2}\ltimes\left(\left(\mathbf{7+1},4\cdot\mathbf{1}+2\cdot\mathbf{3}\right)\times2\cdot\left(\left(\mathbf{7},\mathbf{1}\right)+\left(\mathbf{1},\mathbf{3}\right)\right)\right)},\label{DR3-plane}
\end{equation}
with the desired dimension count
\begin{align}
\text{dim}\mathbb{T}P_{III}^{1} & =14+21+210-14-3-80-20\\
\notag & =128=2\,\,\text{dim}\mathbb{T}.
\end{align}
Here again, the coset (\ref{DR3-plane}) is not symmetric, since
\begin{equation}
\left[\mathfrak{c}\left(\mathbb{T}P_{III}^{2}\right),\mathfrak{c}\left(\mathbb{T}P_{III}^{2}\right)\right]\simeq16\cdot\left(\mathbf{7}+\mathbf{1},\mathbf{3}+\mathbf{1}\right)\otimes_{a}\left(\mathbf{7}+\mathbf{1},\mathbf{3}+\mathbf{1}\right)\nsubseteq\mathfrak{isot}\left(\mathbb{T}P_{III}^{2}\right).
\end{equation}

\begin{center}
\begin{table}
\begin{centering}
\begin{tabular}{|c|c|c|}
\hline
Plane & Covariance & Realization of $\mathbb{R}\otimes\mathbb{C}\otimes\mathbb{H}\otimes\mathbb{O}$\tabularnewline
\hline
\hline
$\mathbb{T}P_{I}^{2}$ & $\mathfrak{der}\left(\mathbb{C}\otimes\mathbb{H}\right)\oplus\mathfrak{der}\left(\mathbb{O}\right)$,
i.e., $\mathfrak{su}_{2}\oplus\mathfrak{g}_{2}$ & $\left(\mathbf{7}+\mathbf{1},\mathbf{7}+\mathbf{1}\right)$\tabularnewline
\hline
$\mathbb{T}P_{II}^{2}$ & $\mathfrak{der}\left(\mathbb{O}\right)\oplus\mathfrak{der}\left(\mathbb{C}\otimes\mathbb{H}\right)$,
i.e., $\mathfrak{g}_{2}\oplus\mathfrak{su}_{2}$ & $\left(\mathbf{7}+\mathbf{1},\mathbf{7}+\mathbf{1}\right)$\tabularnewline
\hline
$\mathbb{T}P_{III}^{2}$ & $\mathfrak{der}\left(\mathbb{C}\otimes\mathbb{O}\right)\oplus\mathfrak{der}\left(\mathbb{H}\right)$,
i.e., $\mathfrak{g}_{2}\oplus\mathfrak{su}_{2}$ & $2\cdot\left(\mathbf{7}+\mathbf{1},\mathbf{3}+\mathbf{1}\right)$\tabularnewline
\hline
\end{tabular}
\par\end{centering}
\centering{}\caption{Realization of the Dixon algebra $\mathbb{R}\otimes\mathbb{C}\otimes\mathbb{H}\otimes\mathbb{O}$
with respect to the maximal manifest covariance of the isometry group
in the three different Dixon-Rosenfeld planes. }
\end{table}
\par\end{center}

\begin{center}
\begin{table}
\begin{centering}
{\footnotesize{}}%
\begin{tabular}{|c|c|c|c|c|}
\hline
{\footnotesize{}Plane} & {\footnotesize{}Isometry Group} & {\footnotesize{}Dim} & {\footnotesize{}Isotropy Group} &
{\footnotesize{}Dim}\tabularnewline
\hline
\hline
{\footnotesize{}$\mathbb{T}P_{I}^{2}$} & {\footnotesize{}$\begin{array}{c}
\left(SU_{2}\times F_{4}\right)\ltimes\\
\left(\boldsymbol{7},\boldsymbol{26}\right)
\end{array}$} & {\footnotesize{}237} & {\footnotesize{}$\begin{array}{c}
\left(SU_{2}\times G_{2}\right)\ltimes\\
\left(\left(\mathbf{7},3\cdot\mathbf{1+7}\right)\times\left(\mathbf{1},3\cdot\mathbf{7+1}\right)\right)
\end{array}$} & {\footnotesize{}109}\tabularnewline
\hline
{\footnotesize{}$\mathbb{T}P_{II}^{2}$} & {\footnotesize{}$\begin{array}{c}
\left(G_{2}\times SU_{2}\right)\ltimes\\
\left(\left(\mathbf{7},3\cdot\mathbf{7+}5\cdot\mathbf{1}\right)\times2\cdot\left(\mathbf{1},\mathbf{7+1}\right)\right)
\end{array}$} & {\footnotesize{}215} & {\footnotesize{}$\begin{array}{c}
\left(G_{2}\times SU_{2}\right)\ltimes\\
\left(\mathbf{7},\mathbf{7+}3\cdot\mathbf{1}\right)
\end{array}$} & {\footnotesize{}87}\tabularnewline
\hline
{\footnotesize{}$\mathbb{T}P_{III}^{2}$} & {\footnotesize{}$\begin{array}{c}
\left(G_{2}\times Sp_{6}\right)\ltimes\\
\left(2\cdot\mathbf{7}+\mathbf{1},\mathbf{14}\right)
\end{array}$} & {\footnotesize{}245} & {\footnotesize{}$\begin{array}{c}
\left(G_{2}\times SU_{2}\right)\ltimes\\
\left(\left(\mathbf{7+1},4\cdot\mathbf{1}+2\cdot\mathbf{3}\right)\times\right.\\
\left.2\cdot\left(\left(\mathbf{7},\mathbf{1}\right)+\left(\mathbf{1},\mathbf{3}\right)\right)\right)
\end{array}$} & {\footnotesize{}117}\tabularnewline
\hline
\end{tabular}{\footnotesize\par}
\par\end{centering}
\centering{}\caption{The three Dixon-Rosenfeld plane with isometry group and isometry group
with their respective dimensions.}
\end{table}
\par\end{center}

\paragraph*{Relations between Dixon-Rosenfeld planes} \,

We have defined three Dixon-Rosenfeld planes as the coset manifolds
\begin{eqnarray}
\mathbb{T}P_{I}^{2} & \simeq & \frac{SU_{2}\times F_{4}\ltimes\left(\mathbf{7},\mathbf{26}\right)}{SU_{2}\times
G_{2}\ltimes\left(\left(\mathbf{7},3\cdot\mathbf{1+7}\right)\times\left(\mathbf{1},3\cdot\mathbf{7+1}\right)\right)},\\
\mathbb{T}P_{II}^{2} & \simeq & \frac{G_{2}\times
SU_{2}\ltimes\left(\left(\mathbf{7},3\cdot\mathbf{7+}5\cdot\mathbf{1}\right)\times2\cdot\left(\mathbf{1},\mathbf{7+1}\right)\right)}{G_{2}\times
SU_{2}\ltimes\left(\mathbf{7},\mathbf{7+}3\cdot\mathbf{1}\right)},\\
\mathbb{T}P_{III}^{2} & \simeq & \frac{G_{2}\times Sp_{6}\ltimes\left(2\cdot\mathbf{7}+\mathbf{1},\mathbf{14}\right)}{G_{2}\times
SU_{2}\ltimes\left(\left(\mathbf{7+1},4\cdot\mathbf{1}+2\cdot\mathbf{3}\right)\times2\cdot\left(\left(\mathbf{7},\mathbf{1}\right)+\left(\mathbf{1},\mathbf{3}\right)\right)\right)}.
\end{eqnarray}
Since our main interest for physical applications is on the isometry
algebra of the planes, we will now briefly analyze the mutual
relationships between these. By comparing the isometry algebra of
the first plane, i.e.,
\begin{eqnarray}
\mathfrak{isom}\left(\mathbb{T}P_{I}^{2}\right) & \simeq &
\mathfrak{f}_{4}\oplus\mathfrak{su}_{2}\oplus\left(\mathbf{26},\mathbf{7}\right)\nonumber \\
 & = & \mathfrak{so}_{9}\oplus\mathfrak{su}_{2}\oplus\left(\mathbf{16+9+1},\mathbf{7}\right)\oplus\left(\mathbf{16},\mathbf{1}\right)\\
 & = &
 \mathfrak{so}_{8}\oplus\mathfrak{su}_{2}\oplus\left(\mathbf{8}_{v}+\mathbf{8}_{s}+\mathbf{8}_{c}+2\cdot\mathbf{1},\mathbf{7}\right)\oplus\left(\mathbf{8}_{v}+\mathbf{8}_{s}+\mathbf{8}_{c},\mathbf{1}\right)\nonumber
 \\
 & = &
 \mathfrak{so}_{7}\oplus\mathfrak{su}_{2}\oplus\left(\mathbf{7}+2\cdot\mathbf{8}+3\cdot\mathbf{1},\mathbf{7}\right)\oplus\left(2\cdot\mathbf{7+1}+2\cdot\mathbf{8},\mathbf{1}\right)\nonumber
 \\
 & = &
 \mathfrak{g}_{2}\oplus\mathfrak{su}_{2}\oplus3\cdot\left(\mathbf{7},\mathbf{7}\right)\oplus5\cdot\left(\mathbf{1},\mathbf{7}\right)\oplus5\cdot\left(\mathbf{7},\mathbf{1}\right)\oplus3\cdot\left(\mathbf{1},\mathbf{1}\right),
\end{eqnarray}
with the one of the second plane, i.e., $\mathfrak{isom}\left(\mathbb{T}P_{II}^{2}\right)$
given by $\mathfrak{A}_{2}^{\text{enh}}$ in (\ref{en1-plane}), it
follows that $\mathfrak{isom}\left(\mathbb{T}P_{II}^{2}\right)\subsetneq\mathfrak{isom}\left(\mathbb{T}P_{I}^{2}\right).$ On
the other hand, by comparing (\ref{isot1-plane}) and (\ref{isot2-plane}),
the same relation holds between the isotropy algebras, i.e.,
$\mathfrak{isot}\left(\mathbb{T}P_{II}^{2}\right)\subsetneq\mathfrak{isot}\left(\mathbb{T}P_{I}^{2}\right).$

As for the relations between the first and the third plane, i.e., $\mathbb{T}P_{I}^{2}$
and $\mathbb{T}P_{III}^{2}$, we have
\begin{align}
\mathfrak{isom}\left(\mathbb{T}P_{I}^{2}\right)\simeq & \mathfrak{f}_{4}\oplus\mathfrak{su}_{2}\oplus\left(\mathbf{26},\mathbf{7}\right)\\
= & \mathfrak{sp}_{6}\oplus\mathfrak{su}_{2}\oplus\mathfrak{su}_{2}\oplus\\
 &
 \oplus\left(\mathbf{14},\mathbf{1},\mathbf{7}\right)\oplus\left(\mathbf{6},\mathbf{2},\mathbf{7}\right)\oplus\left(\mathbf{14}^{\prime},\mathbf{2,1}\right),
\end{align}
where we intended $\mathbf{14}^{\prime}\equiv\wedge_{0}^{3}\mathbf{6}\text{~of~}\mathfrak{sp}_{6}.$ Therefore,
from (\ref{isom3-plane}), we have
\begin{equation}
\mathfrak{isom}\left(\mathbb{T}P_{III}^{1}\right)\simeq\mathfrak{sp}_{6}\oplus\mathfrak{g}_{2}\oplus2\cdot\left(\mathbf{14,7}\right)\oplus\left(\mathbf{14,1}\right).
\end{equation}
We immediately notice that the isometry algebra $\mathfrak{isom}\left(\mathbb{T}P_{I}^{2}\right)$
is not contained within and cannot contain $\mathfrak{isom}\left(\mathbb{T}P_{III}^{2}\right)$.
On the other hand, it follows from (\ref{isot1-plane}) and (\ref{isot3-plane}),
that the same happens for the isotropy algebras $\mathfrak{isot}\left(\mathbb{T}P_{I}^{2}\right)$
and $\mathfrak{isot}\left(\mathbb{T}P_{III}^{2}\right)$.

By virtue of (\ref{en1-plane}), (\ref{isot2-plane}), (\ref{isom3-plane})
and (\ref{isot3-plane}), one obtains that the same results hold
in the case of $\mathbb{T}P_{II}^{2}$ such that the isometry algebra
$\mathfrak{isom}\left(\mathbb{T}P_{II}^{2}\right)$ is not contained within
and cannot contain $\mathfrak{isom}\left(\mathbb{T}P_{III}^{2}\right)$.

\section{\noun{Relations with the Rosenfeld planes}}

It is interesting to point out the relations between the isometry
algebras of the Dixon-Rosenfeld planes and those of the octonionic
Rosenfeld planes, the exceptional Lie algebras $\mathfrak{f}_{4},\mathfrak{e}_{6},\mathfrak{e}_{7}$
and $\mathfrak{e}_{8}$. The definitions of the octonionic Rosenfeld
planes from a historical perspective are found in \cite{Rosenfeld Group2-1,Rosenf98}, with
a more rigorous approach found in \cite{Lands-Maniv}.
In another paper, we presented an alternative definition of all the possible Rosenfeld planes using a magic square approach \cite{Corr Rosen}.
Let us recall here the homogeneous space realizations of the
Rosenfeld planes over $\mathbb{A}\otimes\mathbb{O}$, with\footnote{Note that
$\mathfrak{isot}\left(\left(\mathbb{A}\otimes\mathbb{O}\right)P^{2}\right)\simeq\mathfrak{isom}\left(\left(\mathbb{A}\otimes\mathbb{O}\right)P^{1}\right)$
(including in the latter the global algebras
$\mathcal{A}_{q}\simeq\mathfrak{tri}\left(\mathbb{A}\right)\ominus\mathfrak{soi}\left(\mathbb{A}\right)=\varnothing,\mathfrak{u}_{1},\mathfrak{su}_{2},\varnothing$
for $q=\dim_{\mathbb{R}}\mathbb{A}=1,2,4,8$ - corresponding to $\mathbb{A}=\mathbb{R},\mathbb{C},\mathbb{H},\mathbb{O}$
-, respectively).} $\mathbb{A}=\mathbb{R},\mathbb{C},\mathbb{H},\mathbb{O}$ (see \cite{Rosenf98,Rosenfeld Group2-1}
and \cite{Corr Rosen,Corr Notes Octo,RealF}), i.e., for the \emph{octonionic
projective plane} $\left(\mathbb{R}\otimes\mathbb{O}\right)P^{2}$,
the \emph{bioctonionic Rosenfeld plane} $\left(\mathbb{C}\otimes\mathbb{O}\right)P^{2}$,
the \emph{quateroctonionic Rosenfeld plane} $\left(\mathbb{H}\otimes\mathbb{O}\right)P^{2}$
and, finally, for the \emph{octooctonionic Rosenfeld plane} $\left(\mathbb{O}\otimes\mathbb{O}\right)P^{2}$:
\begin{align*}
\left(\mathbb{R}\otimes\mathbb{O}\right)P^{2} & =\frac{F_{4}}{SO_{9}},\\
\left(\mathbb{C}\otimes\mathbb{O}\right)P^{2} & =\frac{E_{6}}{SO_{10}\times U_{1}},\\
\left(\mathbb{H}\otimes\mathbb{O}\right)P^{2} & =\frac{E_{7}}{SO_{12}\times SU_{2}},\\
\left(\mathbb{O}\otimes\mathbb{O}\right)P^{2} & =\frac{E_{8}}{SO_{16}}.
\end{align*}
From such realizations, it follows that\footnote{The representations of the tangent spaces are up to \textit{spinor
polarizations} \cite{Minchenko}.} the tangent space $T\left(\mathbb{O}P^{2}\right)$ can be identified
with the $\mathbf{16}$ representation of $\mathfrak{so}_{9},$ and
since it is also equivalent to $2\cdot\mathbf{8}$ of $\mathfrak{so}_{7}$, we then have
\begin{equation}
T\left(\mathbb{O}P^{2}\right)\simeq2\cdot\left(\mathbf{7}+\mathbf{1}\right)~\text{of~}\mathfrak{g}_{2}.
\end{equation}
Similarly, in the \emph{bioctonionic} case, we have that $T\left(\left(\mathbb{C}\otimes\mathbb{O}\right)P^{2}\right)$
can be identified with the $\mathbf{16}_{+}\oplus\overline{\mathbf{16}}_{-}$
representation of $\mathfrak{so}_{10}\oplus\mathfrak{u}_{1}$, which
can be thought of as $4\cdot\mathbf{8}~\text{of~}\mathfrak{so}_{7}$
and therefore,
\begin{equation}
T\left(\left(\mathbb{C}\otimes\mathbb{O}\right)P^{2}\right)\simeq4\cdot\left(\mathbf{7}+\mathbf{1}\right)~\text{of~\ensuremath{\mathfrak{g}_{2}}.}
\end{equation}
For the \emph{quateroctonionic} Rosenfeld plane, the identification of
$T\left(\left(\mathbb{H}\otimes\mathbb{O}\right)P^{2}\right)$ with
the $\left(\mathbf{32},\mathbf{2}\right)$ of $\mathfrak{so}_{12}\oplus\mathfrak{su}_{2}$
gives
\begin{equation}
T\left(\left(\mathbb{H}\otimes\mathbb{O}\right)P^{2}\right)\simeq2\cdot\left(\mathbf{7+1},\mathbf{3}+\mathbf{1}\right)~\text{of~}\mathfrak{g}_{2}\oplus\mathfrak{su}_{2}.
\end{equation}
Finally, in the \emph{octooctonic} case, the identification of $T\left(\left(\mathbb{O}\otimes\mathbb{O}\right)P^{2}\right)$
with the $\mathbf{128}~\text{of~}\mathfrak{so}_{16}$ can be thought of
as $2\cdot\left(\mathbf{8},\mathbf{8}\right)~\text{of~}\mathfrak{so}_{7}\oplus\mathfrak{so}_{7}$,
therefore giving
\[
T\left(\left(\mathbb{O}\otimes\mathbb{O}\right)P^{2}\right)\simeq2\cdot\left(\mathbf{7+1},\mathbf{7+1}\right)~\text{of~}\mathfrak{g}_{2}\oplus\mathfrak{g}_{2}.
\]
Considering that $\mathfrak{g}_{2}=\mathfrak{der}\left(\mathbb{O}\right)=\mathfrak{der}\left(\mathbb{C}\otimes\mathbb{O}\right)$,
$\mathfrak{g}_{2}\oplus\mathfrak{su}_{2}=\mathfrak{der}\left(\mathbb{H}\otimes\mathbb{O}\right)$,
and $\mathfrak{g}_{2}\oplus\mathfrak{g}_{2}=\mathfrak{der}\left(\mathbb{O}\otimes\mathbb{O}\right)$,
the previous relations illustrate how the tangent spaces of octonionic
projective planes generally carry an enhancement of the symmetry with
respect to the Lie algebra
$\mathfrak{der}\left(\mathbb{A}\otimes\mathbb{O}\right)\simeq\mathfrak{der}\left(\mathbb{A}\right)\oplus\mathfrak{g}_{2}$.
Furthermore, from the previous characterization, we also have
\begin{eqnarray}
\text{dim}\,\mathbb{O}P^{2} & = & 16,\\
\text{dim\,}\left(\mathbb{C}\otimes\mathbb{O}\right)P^{2} & = & 32,\\
\text{dim\,}\left(\mathbb{H}\otimes\mathbb{O}\right)P^{2} & = & 64,\\
\text{dim\,}\left(\mathbb{O}\otimes\mathbb{O}\right)P^{2} & = & 128.
\end{eqnarray}

We now have all the ingredients to see the relationships between Dixon-Rosenfeld
planes and the octonionic Rosenfeld planes. Of course,
\begin{equation}
\text{dim}\left(\mathbb{O}\otimes\mathbb{O}\right)P^{2}=\text{dim}\left(\mathbb{T}P_{I}^{2}\right)=\text{dim}\left(\mathbb{T}P_{II}^{2}\right)=\text{dim}\left(\mathbb{T}P_{III}^{2}\right)=128,
\end{equation}
but the relations among such projective spaces are generally not trivial,
as resulting from the discussion below.

\paragraph*{Relations with $\mathbb{T}P_{I}^{2}$} \,

By recalling (\ref{TI-plane}), we report here the $\left(\mathfrak{g}_{2}\oplus\mathfrak{g}_{2}\right)$-covariant
representation of two copies of the tensor algebra
\begin{equation}
\left(\mathbb{O}\otimes\mathbb{O}\right)^{\oplus2}\simeq2\cdot\left(\mathbf{7}+\mathbf{1},\mathbf{7}+\mathbf{1}\right)~\text{of~}\mathfrak{g}_{2}\oplus\mathfrak{g}_{2},
\end{equation}
where $\mathfrak{g}_{2}\oplus\mathfrak{g}_{2}$ is $\mathfrak{der}\left(\mathbb{O}\right)\oplus\mathfrak{der}\left(\mathbb{O}\right)$.
Again, one observes that, when restricting $\mathfrak{g}_{2}\rightarrow\mathfrak{su}_{2,P}$
(where the subscript ``$P$''\ denotes the \textit{principal} $\mathfrak{su}_{2}$ \cite{Kostant}), the irrepr. $\mathbf{7}$ of $\mathfrak{g}_{2}$
remains irreducible into the spin-3 irrepr. $\mathbf{7}$ of $\mathfrak{su}_{2}$,
and therefore, the passage from the octooctonions to the Dixon algebra
is characterized by
\begin{equation}
\mathbb{O}\otimes\mathbb{O}\simeq\underset{\mathfrak{g}_{2}\oplus\mathfrak{g}_{2}}{\left(\mathbf{7}+\mathbf{1},\mathbf{7}+\mathbf{1}\right)}\overset{\mathfrak{g}_{2}\rightarrow\mathfrak{su}_{2,P}}{\simeq}\underset{\mathfrak{su}_{2}\oplus\mathfrak{g}_{2}}{\left(\mathbf{7}+\mathbf{1},\mathbf{7}+\mathbf{1}\right)}\simeq\mathbb{T}.
\end{equation}

On the other hand, since the isometry algebra of the octooctonionic
Rosenfeld plane $\left(\mathbb{O}\otimes\mathbb{O}\right)P^{2}$ is
given by
\begin{eqnarray}
\mathfrak{isom}\left(\left(\mathbb{O}\otimes\mathbb{O}\right)P^{2}\right) & \simeq & \mathfrak{e}_{8}\nonumber \\
\notag & = & \mathfrak{f}_{4}\oplus\mathfrak{g}_{2}\oplus\left(\mathbf{26},\mathbf{7}\right)\\
\notag & = & \mathfrak{f}_{4}\oplus\mathfrak{su}_{2,P}\oplus\left(\mathbf{26},\mathbf{7}\right)\oplus\left(\mathbf{1},\mathbf{11}\right)\\
 & \simeq & \mathfrak{isom}\left(\mathbb{T}P_{I}^{2}\right)\oplus\left(\mathbf{1},\mathbf{11}\right),
\end{eqnarray}
and the isotropy algebra
\begin{eqnarray}
\mathfrak{isot}\left(\left(\mathbb{O}\otimes\mathbb{O}\right)P^{2}\right) & \simeq & \mathfrak{so}_{16}\nonumber \\
\notag & = & \mathfrak{so}_{8}\oplus\mathfrak{so}_{8}\oplus\left(\mathbf{8}_{v},\mathbf{8}_{v}\right)\\
\notag & = &
\mathfrak{so}_{7}\oplus\mathfrak{so}_{7}\oplus\left(\mathbf{7+1},\mathbf{7+1}\right)\oplus\left(\mathbf{7},\mathbf{1}\right)\oplus\left(\mathbf{1},\mathbf{7}\right)\\
 & = &
 \mathfrak{g}_{2}\oplus\mathfrak{g}_{2}\oplus\left(\mathbf{7+1},\mathbf{7+1}\right)\oplus2\cdot\left(\mathbf{7},\mathbf{1}\right)\oplus2\cdot\left(\mathbf{1},\mathbf{7}\right)\nonumber
 \\
 & = &
 \mathfrak{g}_{2}\oplus\mathfrak{su}_{2}\oplus\left(\mathbf{7+1},\mathbf{7+1}\right)\oplus2\cdot\left(\mathbf{7},\mathbf{1}\right)\oplus2\cdot\left(\mathbf{1},\mathbf{7}\right)\oplus\left(\mathbf{1},\mathbf{11}\right)\nonumber
 \\
 & \simeq & \mathfrak{isot}\left(\mathbb{T}P_{I}^{2}\right)\oplus\left(\mathbf{1},\mathbf{11}\right),
\end{eqnarray}
it holds that
\begin{eqnarray}
\mathfrak{isom}\left(\mathbb{T}P_{I}^{2}\right) & \subsetneq & \mathfrak{isom}\left(\left(\mathbb{O}\otimes\mathbb{O}\right)P^{2}\right),\\
\mathfrak{isot}\left(\mathbb{T}P_{I}^{2}\right) & \subsetneq & \mathfrak{isot}\left(\left(\mathbb{O}\otimes\mathbb{O}\right)P^{2}\right).
\end{eqnarray}

{\rr Notice that the isometry and isotropy Lie algebras of $\left( \mathbb{O}%
\otimes \mathbb{O}\right) P^{2}$ is enlarged from the isometry
and isotropy Lie algebras of $\mathbb{T}P_{I}^{2}$ by the following $\mathfrak{su}_{2}$-covariant set of generators, respectively:
\begin{equation}
\mathfrak{isom}\left( \left( \mathbb{O}\otimes \mathbb{O}\right)
P^{2}\right) \ominus \mathfrak{isom}\left( \mathbb{T}P_{I}^{2}\right) \simeq
\mathfrak{isot}\left( \left( \mathbb{O}\otimes \mathbb{O}\right)
P^{2}\right) \ominus \mathfrak{isot}\left( \mathbb{T}P_{I}^{2}\right) \simeq
\mathbf{11},  \label{enn-plane}
\end{equation}%
i.e., by the spin-5 irrepr. $\mathbf{11}$ of $\mathfrak{su}_{2}$.}

Passing to the quateroctonionic Rosenfeld plane, we notice that since
\begin{eqnarray}
\notag\mathfrak{isom}\left(\left(\mathbb{H}\otimes\mathbb{O}\right)P^{2}\right) & \simeq & \mathfrak{e}_{7}\\
 & = & \mathfrak{f}_{4}\oplus\mathfrak{su}_{2}\oplus\left(\mathbf{26},\mathbf{3}\right),\\
\mathfrak{isot}\left(\left(\mathbb{H}\otimes\mathbb{O}\right)P^{2}\right) & \simeq & \mathfrak{so}_{12}\oplus\mathfrak{su}_{2}\nonumber \\
\notag & = & \mathfrak{so}_{7}\oplus\mathfrak{so}_{5}\oplus\mathfrak{su}_{2}\oplus\left(\mathbf{7},\mathbf{5,1}\right)\\
 & = &
 \mathfrak{g}_{2}\oplus\mathfrak{so}_{5}\oplus\mathfrak{su}_{2}\oplus\left(\mathbf{7},\mathbf{5,1}\right)\oplus\left(\mathbf{7},\mathbf{1,1}\right),
\end{eqnarray}
and since (\ref{isot1-plane}), we then have that
\begin{eqnarray}
\mathfrak{isom}\left(\mathbb{T}P_{I}^{2}\right) & \nsubseteq &
\nsupseteq\mathfrak{isom}\left(\left(\mathbb{H}\otimes\mathbb{O}\right)P^{2}\right),\\
\mathfrak{isot}\left(\mathbb{T}P_{I}^{2}\right) & \nsubseteq &
\nsupseteq\mathfrak{isot}\left(\left(\mathbb{H}\otimes\mathbb{O}\right)P^{2}\right).
\end{eqnarray}

Finally, in the case of the bioctonionic Rosenfeld plane, the isometry
and isotropy algebras are given by
\begin{eqnarray}
\notag\mathfrak{isom}\left(\left(\mathbb{C}\otimes\mathbb{O}\right)P^{2}\right) & \simeq & \mathfrak{e}_{6}\\
 & = & \mathfrak{f}_{4}\oplus\mathbf{26},\\
\mathfrak{isot}\left(\left(\mathbb{C}\otimes\mathbb{O}\right)P^{2}\right) & \simeq & \mathfrak{so}_{10}\oplus\mathfrak{u}_{1}\nonumber \\
 & = & \mathfrak{so}_{7}\oplus\mathfrak{su}_{2}\oplus\mathfrak{u}_{1}\oplus\left(\mathbf{7},\mathbf{3}\right)_{0}\\
 & = &
 \mathfrak{g}_{2}\oplus\mathfrak{su}_{2}\oplus\mathfrak{u}_{1}\oplus\left(\mathbf{7},\mathbf{3}\right)_{0}\oplus\left(\mathbf{7},\mathbf{1}\right)_{0}\nonumber
 \\
 & = &
 \mathfrak{g}_{2}\oplus\mathfrak{su}_{2}\oplus\left(\mathbf{7},\mathbf{3}\right)\oplus\left(\mathbf{7},\mathbf{1}\right)\oplus\left(\mathbf{1},\mathbf{1}\right).
\end{eqnarray}
By recalling (\ref{TI-plane}), the (doubling of the) $\mathfrak{g}_{2}$-covariant
representation of two copies of the tensor algebra $\mathbb{C}\otimes\mathbb{O}$
and (\ref{isot1-plane}), we then have
\begin{eqnarray}
\mathfrak{isom}\left(\left(\mathbb{C}\otimes\mathbb{O}\right)P^{2}\right) & \subsetneq & \mathfrak{isom}\left(\mathbb{T}P_{I}^{2}\right),\\
\mathfrak{isot}\left(\left(\mathbb{C}\otimes\mathbb{O}\right)P^{2}\right) & \nsubseteq &
\nsupseteq\mathfrak{isot}\left(\mathbb{T}P_{I}^{2}\right).
\end{eqnarray}

\begin{center}
\begin{table}
\begin{centering}
{\footnotesize{}}%
\begin{tabular}{|c|c|}
\hline
{\footnotesize{}Isometry Algebra of the Plane} & {\footnotesize{}Relations with Isometry Algebra of the Rosenfeld Planes}\tabularnewline
\hline
\hline
{\footnotesize{}$\mathfrak{isom}\left(\mathbb{T}P_{I}^{2}\right)$} &
{\footnotesize{}$\mathfrak{isom}\left(\mathbb{O}P^{2}\right)\subset\mathfrak{isom}\left(\left(\mathbb{C}\otimes\mathbb{O}\right)P^{2}\right)\subset\mathfrak{isom}\left(\mathbb{T}P_{I}^{2}\right)\subset\mathfrak{isom}\left(\left(\mathbb{O}\otimes\mathbb{O}\right)P^{2}\right)$}\tabularnewline
\hline
{\footnotesize{}$\mathfrak{isom}\left(\mathbb{T}P_{II}^{2}\right)$} &
{\footnotesize{}$\mathfrak{isom}\left(\mathbb{T}P_{II}^{2}\right)\subset\mathfrak{isom}\left(\left(\mathbb{O}\otimes\mathbb{O}\right)P^{2}\right)$}\tabularnewline
\hline
{\footnotesize{}$\mathfrak{isom}\left(\mathbb{T}P_{III}^{2}\right)$} &
{\footnotesize{}$\mathfrak{isom}\left(\mathbb{O}P^{2}\right)\subset\mathfrak{isom}\left(\left(\mathbb{C}\otimes\mathbb{O}\right)P^{2}\right)\subset\mathfrak{isom}\left(\left(\mathbb{H}\otimes\mathbb{O}\right)P^{2}\right)\subset\mathfrak{isom}\left(\mathbb{T}P_{III}^{2}\right)$}\tabularnewline
\hline
\end{tabular}{\footnotesize\par}
\par\end{centering}
\medskip{}

\begin{centering}
{\footnotesize{}}%
\begin{tabular}{|c|c|}
\hline
{\footnotesize{}Isotropy Algebra of the Plane} & {\footnotesize{}Relations with Isometry Algebra of the Rosenfeld Planes}\tabularnewline
\hline
\hline
{\footnotesize{}$\mathfrak{isot}\left(\mathbb{T}P_{I}^{2}\right)$} &
{\footnotesize{}$\mathfrak{isot}\left(\mathbb{T}P_{I}^{2}\right)\subset\mathfrak{isot}\left(\left(\mathbb{O}\otimes\mathbb{O}\right)P^{2}\right)$}\tabularnewline
\hline
{\footnotesize{}$\mathfrak{isot}\left(\mathbb{T}P_{II}^{2}\right)$} &
{\footnotesize{}$\mathfrak{isot}\left(\mathbb{T}P_{II}^{2}\right)\subset\mathfrak{isot}\left(\left(\mathbb{O}\otimes\mathbb{O}\right)P^{2}\right)$}\tabularnewline
\hline
{\footnotesize{}$\mathfrak{isot}\left(\mathbb{T}P_{III}^{2}\right)$} &
{\footnotesize{}$\mathfrak{isot}\left(\mathbb{O}P^{2}\right)\subset\mathfrak{isot}\left(\left(\mathbb{C}\otimes\mathbb{O}\right)P^{2}\right)\subset\mathfrak{isot}\left(\left(\mathbb{H}\otimes\mathbb{O}\right)P^{2}\right)\subset\mathfrak{isot}\left(\mathbb{T}P_{III}^{2}\right)$}\tabularnewline
\hline
\end{tabular}{\footnotesize\par}
\par\end{centering}
\centering{}\caption{\label{tab:Relations The-three-Dixon-Rosenfeld-1}The isometry {\rr and isotropy} algebras
of the three Dixon-Rosenfeld plane, in their relationships with the
isometry and isotropy algebras of the octonionic Rosenfeld planes.}
\end{table}
\par\end{center}

\paragraph*{Relations with $\mathbb{T}P_{II}^{2}$} \,

The same line of reasoning on the algebra level leads to the inclusion
of the second Dixon-Rosenfeld plane $\mathbb{T}P_{II}^{2}$ within the
octooctonionic plane $\left(\mathbb{O}\otimes\mathbb{O}\right)P^{2}$.
By iterated branchings, one computes that
\begin{eqnarray}
\mathfrak{isom}\left(\left(\mathbb{O}\otimes\mathbb{O}\right)P^{2}\right) & \simeq & \mathfrak{e}_{8}\nonumber \\
 & = & \mathfrak{f}_{4}\oplus\mathfrak{g}_{2}\oplus\left(\mathbf{26},\mathbf{7}\right)\\
 & = &
 \mathfrak{g}_{2}\oplus\mathfrak{su}_{2}\oplus\mathfrak{g}_{2}\oplus\left(\mathbf{7},\mathbf{3,7}\right)\oplus\left(\mathbf{1},\mathbf{5,7}\right)\oplus\left(\mathbf{7},\mathbf{5,1}\right)\nonumber
 \\
 & = &
 \mathfrak{g}_{2}\oplus\mathfrak{su}_{2}\oplus\mathfrak{su}_{2,P}\oplus\left(\mathbf{7},\mathbf{3,7}\right)\oplus\left(\mathbf{1},\mathbf{5,7}\right)\oplus\left(\mathbf{7},\mathbf{5,1}\right)\oplus\left(\mathbf{1},\mathbf{1,11}\right)\nonumber
 \\
 & = &
 \mathfrak{g}_{2}\oplus\mathfrak{su}_{2,P}\oplus3\cdot\left(\mathbf{7},\mathbf{7}\right)\oplus5\cdot\left(\mathbf{7},\mathbf{1}\right)\oplus5\cdot\left(\mathbf{1},\mathbf{7}\right)\oplus3\cdot\left(\mathbf{1},\mathbf{1}\right)\oplus\left(\mathbf{1},\mathbf{11}\right)\nonumber
 \\
 & = &
 \mathfrak{isom}\left(\mathbb{T}P_{II}^{2}\right)\oplus3\cdot\left(\mathbf{1},\mathbf{7}\right)\oplus\left(\mathbf{1},\mathbf{1}\right)\oplus\left(\mathbf{1},\mathbf{11}\right),
\end{eqnarray}
and, as for the isotropy algebra we have
\begin{eqnarray}
\mathfrak{isot}\left(\left(\mathbb{O}\otimes\mathbb{O}\right)P^{2}\right) & \simeq & \mathfrak{so}_{16}\nonumber \\
 & = & \mathfrak{so}_{8}\oplus\mathfrak{so}_{8}\oplus\left(\mathbf{8}_{v},\mathbf{8}_{v}\right)\\
 & = &
 \mathfrak{so}_{7}\oplus\mathfrak{so}_{7}\oplus\left(\mathbf{7+1},\mathbf{7+1}\right)\oplus\left(\mathbf{7},\mathbf{1}\right)\oplus\left(\mathbf{1},\mathbf{7}\right)\nonumber
 \\
 & = &
 \mathfrak{g}_{2}\oplus\mathfrak{g}_{2}\oplus\left(\mathbf{7+1},\mathbf{7+1}\right)\oplus2\cdot\left(\mathbf{7},\mathbf{1}\right)\oplus2\cdot\left(\mathbf{1},\mathbf{7}\right)\nonumber
 \\
 & = &
 \mathfrak{g}_{2}\oplus\mathfrak{su}_{2}\oplus\left(\mathbf{7},\mathbf{7}\right)\oplus3\cdot\left(\mathbf{7},\mathbf{1}\right)\oplus3\cdot\left(\mathbf{1},\mathbf{7}\right)\oplus\left(\mathbf{1},\mathbf{1}\right)\oplus\left(\mathbf{1},\mathbf{11}\right)\nonumber
 \\
 & \simeq &
 \mathfrak{isot}\left(\mathbb{T}P_{II}^{2}\right)\oplus3\cdot\left(\mathbf{1},\mathbf{7}\right)\oplus\left(\mathbf{1},\mathbf{1}\right)\oplus\left(\mathbf{1},\mathbf{11}\right).
\end{eqnarray}
In other words,
\begin{eqnarray}
\mathfrak{isom}\left(\mathbb{T}P_{II}^{2}\right) & \subsetneq & \mathfrak{isom}\left(\left(\mathbb{O}\otimes\mathbb{O}\right)P^{2}\right),\\
\mathfrak{isot}\left(\mathbb{T}P_{II}^{2}\right) & \subsetneq & \mathfrak{isot}\left(\left(\mathbb{O}\otimes\mathbb{O}\right)P^{2}\right).
\end{eqnarray}
It is interesting to notice that the isometry (and isotropy, resp.) Lie
algebra of $\left(\mathbb{O}\otimes\mathbb{O}\right)P^{2}$ is enlarged
with respect to the isometry (and isotropy, resp.) Lie algebra of $\mathbb{T}P_{II}^{2}$
by the following $\mathfrak{su}_{2}$-covariant set of generators,
i.e.,
\begin{eqnarray}
 &  &
 \mathfrak{isom}\left(\left(\mathbb{O}\otimes\mathbb{O}\right)P^{2}\right)\ominus\mathfrak{isom}\left(\mathbb{T}P_{II}^{2}\right)\simeq\nonumber
 \\
 &  &
 \mathfrak{isot}\left(\left(\mathbb{O}\otimes\mathbb{O}\right)P^{2}\right)\ominus\mathfrak{isot}\left(\mathbb{T}P_{II}^{2}\right)\simeq3\cdot\mathbf{7}\oplus\mathbf{1}\oplus\mathbf{11},\label{enn-plane-2}
\end{eqnarray}
namely by three copies of spin-3, one spin-0 and one spin-5 irreprs.
of $\mathfrak{su}_{2}$.

On the other hand, with the quateroctonionic plane, have that
\begin{eqnarray}
\mathfrak{isom}\left(\left(\mathbb{H}\otimes\mathbb{O}\right)P^{2}\right) & \simeq & \mathfrak{e}_{7}\nonumber \\
 & = & \mathfrak{f}_{4}\oplus\mathfrak{su}_{2}\oplus\left(\mathbf{26},\mathbf{3}\right)\\
 & = &
 \mathfrak{g}_{2}\oplus\mathfrak{su}_{2}\oplus\mathfrak{su}_{2}\oplus\left(\mathbf{7},\mathbf{5,1}\right)\oplus\left(\mathbf{7,3},\mathbf{3}\right)\oplus\left(\mathbf{1,5},\mathbf{3}\right),\\
\mathfrak{isot}\left(\left(\mathbb{H}\otimes\mathbb{O}\right)P^{2}\right) & \simeq & \mathfrak{so}_{12}\oplus\mathfrak{su}_{2}\nonumber \\
 & = & \mathfrak{so}_{7}\oplus\mathfrak{so}_{5}\oplus\mathfrak{su}_{2}\oplus\left(\mathbf{7},\mathbf{5,1}\right)\\
 & = &
 \mathfrak{g}_{2}\oplus\mathfrak{so}_{5}\oplus\mathfrak{su}_{2}\oplus\left(\mathbf{7},\mathbf{5,1}\right)\oplus\left(\mathbf{7},\mathbf{1,1}\right),
\end{eqnarray}
and, since the (\ref{isot1-plane}), we therefore have again the case
of
\begin{eqnarray}
\mathfrak{isom}\left(\mathbb{T}P_{II}^{2}\right) & \nsubseteq &
\nsupseteq\mathfrak{isom}\left(\left(\mathbb{H}\otimes\mathbb{O}\right)P^{2}\right),\\
\mathfrak{isot}\left(\mathbb{T}P_{II}^{2}\right) & \nsubseteq &
\nsupseteq\mathfrak{isot}\left(\left(\mathbb{H}\otimes\mathbb{O}\right)P^{2}\right).
\end{eqnarray}

Finally, for the bioctonionic Rosenfeld plane,
\begin{eqnarray}
\mathfrak{isom}\left(\left(\mathbb{C}\otimes\mathbb{O}\right)P^{2}\right) & \simeq & \mathfrak{e}_{6}\nonumber \\
 & = & \mathfrak{so}_{10}\oplus\mathfrak{u}_{1}\oplus\mathbf{16}_{+}\oplus\overline{\mathbf{16}}_{-}\\
 & = &
 \mathfrak{so}_{7}\oplus\mathfrak{su}_{2}\oplus\mathfrak{u}_{1}\oplus\left(\mathbf{7,3}\right)_{0}\oplus\left(\mathbf{8,2}\right)_{+}\oplus\left(\mathbf{8,2}\right)_{-}\nonumber
 \\
 & = &
 \mathfrak{g}_{2}\oplus\mathfrak{su}_{2}\oplus\mathfrak{u}_{1}\oplus\left(\mathbf{7,3}\right)_{0}\oplus\left(\mathbf{7+1,2}\right)_{+}\oplus\left(\mathbf{7+1,2}\right)_{-}\oplus\left(\mathbf{7,1}\right)_{0}\nonumber
 \\
 & = &
 \mathfrak{g}_{2}\oplus\mathfrak{su}_{2}\oplus\left(\mathbf{7,3}\right)\oplus2\cdot\left(\mathbf{7+1,2}\right)\oplus\left(\mathbf{7,1}\right)\oplus\left(\mathbf{1,1}\right);\\
\mathfrak{isot}\left(\left(\mathbb{C}\otimes\mathbb{O}\right)P^{2}\right) & \simeq & \mathfrak{so}_{10}\oplus\mathfrak{u}_{1}\nonumber \\
 & = & \mathfrak{so}_{7}\oplus\mathfrak{su}_{2}\oplus\mathfrak{u}_{1}\oplus\left(\mathbf{7},\mathbf{3}\right)_{0}\\
 & = &
 \mathfrak{g}_{2}\oplus\mathfrak{su}_{2}\oplus\mathfrak{u}_{1}\oplus\left(\mathbf{7},\mathbf{3}\right)_{0}\oplus\left(\mathbf{7},\mathbf{1}\right)_{0}\nonumber
 \\
 & = &
 \mathfrak{g}_{2}\oplus\mathfrak{su}_{2}\oplus\left(\mathbf{7},\mathbf{3}\right)\oplus\left(\mathbf{7},\mathbf{1}\right)\oplus\left(\mathbf{1},\mathbf{1}\right).
\end{eqnarray}
Recalling (\ref{TI-plane}) and the $\mathfrak{g}_{2}$-covariant
representation of two copies of $\mathbb{C}\otimes\mathbb{O}$ in (\ref{en1-plane})
and (\ref{isot2-plane}), we also have
\begin{eqnarray}
\mathfrak{isom}\left(\left(\mathbb{C}\otimes\mathbb{O}\right)P^{2}\right) & \subsetneq & \mathfrak{isom}\left(\mathbb{T}P_{II}^{2}\right),\\
\mathfrak{isot}\left(\left(\mathbb{C}\otimes\mathbb{O}\right)P^{2}\right) & \nsubseteq &
\nsupseteq\mathfrak{isot}\left(\mathbb{T}P_{II}^{2}\right).
\end{eqnarray}

\paragraph*{Relations with $\mathbb{T}P_{III}^{2}$} \,

The isometries of the third Dixon-Rosenfeld plane are quite different
from the others, indeed by suitably branching $\mathfrak{isom}\left(\left(\mathbb{O}\otimes\mathbb{O}\right)P^{2}\right)$
and $\mathfrak{isot}\left(\left(\mathbb{O}\otimes\mathbb{O}\right)P^{2}\right)$,
one obtains
\begin{eqnarray}
\mathfrak{isom}\left(\left(\mathbb{O}\otimes\mathbb{O}\right)P^{2}\right) & \simeq & \mathfrak{e}_{8}\nonumber \\
 & = & \mathfrak{g}_{2}\oplus\mathfrak{f}_{4}\oplus\left(\mathbf{7},\mathbf{26}\right)\\
 & = &
 \mathfrak{g}_{2}\oplus\mathfrak{sp}_{6}\oplus\mathfrak{su}_{2}\oplus\left(\mathbf{1},\mathbf{14}^{\prime},\mathbf{2}\right)\oplus\left(\mathbf{7},\mathbf{14,1}\right)\oplus\left(\mathbf{7},\mathbf{6,2}\right)\nonumber
 \\
 & = &
 \mathfrak{g}_{2}\oplus\mathfrak{sp}_{6}\oplus2\cdot\left(\mathbf{1},\mathbf{14}^{\prime}\right)\oplus\left(\mathbf{7},\mathbf{14}\right)\oplus2\cdot\left(\mathbf{7},\mathbf{6}\right)\oplus3\cdot\left(\mathbf{1},\mathbf{1}\right).
\end{eqnarray}
On the other hand,
\begin{eqnarray}
\mathfrak{isot}\left(\left(\mathbb{O}\otimes\mathbb{O}\right)P^{2}\right) & \simeq &
\mathfrak{so}_{16}=\mathfrak{so}_{8}\oplus\mathfrak{so}_{8}\oplus\left(\mathbf{8}_{v},\mathbf{8}_{v}\right)\nonumber \\
 & = &
 \mathfrak{so}_{7}\oplus\mathfrak{so}_{7}\oplus\left(\mathbf{7+1},\mathbf{7+1}\right)\oplus\left(\mathbf{7},\mathbf{1}\right)\oplus\left(\mathbf{1},\mathbf{7}\right)\nonumber
 \\
 & = &
 \mathfrak{g}_{2}\oplus\mathfrak{g}_{2}\oplus\left(\mathbf{7+1},\mathbf{7+1}\right)\oplus2\cdot\left(\mathbf{7},\mathbf{1}\right)\oplus2\cdot\left(\mathbf{1},\mathbf{7}\right)\nonumber
 \\
 & = &
 \mathfrak{g}_{2}\oplus\mathfrak{su}_{2,P}\oplus\left(\mathbf{7+1},\mathbf{7+1}\right)\oplus2\cdot\left(\mathbf{7},\mathbf{1}\right)\oplus2\cdot\left(\mathbf{1},\mathbf{7}\right)\oplus(\mathbf{1},\mathbf{11}).
\end{eqnarray}
Thus, by recalling (\ref{isom3-plane}) and (\ref{isot3-plane}),
one obtains
\begin{eqnarray}
\mathfrak{isom}\left(\mathbb{T}P_{III}^{2}\right) & \nsubseteq &
\nsupseteq\mathfrak{isom}\left(\left(\mathbb{O}\otimes\mathbb{O}\right)P^{2}\right),\\
\mathfrak{isot}\left(\mathbb{T}P_{III}^{2}\right) & \nsubseteq &
\nsupseteq\mathfrak{isot}\left(\left(\mathbb{O}\otimes\mathbb{O}\right)P^{2}\right).
\end{eqnarray}
As for the quateroctonionic plane $\left(\mathbb{H}\otimes\mathbb{O}\right)P^{2}$,
we have
\begin{eqnarray}
\mathfrak{isom}\left(\left(\mathbb{H}\otimes\mathbb{O}\right)P^{2}\right) & \simeq & \mathfrak{e}_{7}\nonumber \\
 & = & \mathfrak{f}_{4}\oplus\mathfrak{su}_{2}\oplus\left(\mathbf{26},\mathbf{3}\right)\\
 & = &
 \mathfrak{g}_{2}\oplus\mathfrak{su}_{2}\oplus\mathfrak{su}_{2}\oplus\left(\mathbf{7},\mathbf{5,1}\right)\oplus\left(\mathbf{7,3},\mathbf{3}\right)\oplus\left(\mathbf{1,5},\mathbf{3}\right),\\
\mathfrak{isot}\left(\left(\mathbb{H}\otimes\mathbb{O}\right)P^{2}\right) & \simeq & \mathfrak{so}_{12}\oplus\mathfrak{su}_{2}\nonumber \\
 & = & \mathfrak{so}_{7}\oplus\mathfrak{so}_{5}\oplus\mathfrak{su}_{2}\oplus\left(\mathbf{7},\mathbf{5,1}\right)\\
 & = &
 \mathfrak{g}_{2}\oplus\mathfrak{so}_{5}\oplus\mathfrak{su}_{2}\oplus\left(\mathbf{7},\mathbf{5,1}\right)\oplus\left(\mathbf{7},\mathbf{1,1}\right)\nonumber
 \\
 & = &
 \mathfrak{g}_{2}\oplus\mathfrak{su}_{2}\oplus\mathfrak{su}_{2}\oplus\mathfrak{su}_{2}\oplus\left(\mathbf{7},\mathbf{2,2,1}\right)\oplus2\cdot\left(\mathbf{7},\mathbf{1,1,1}\right)\oplus\left(\mathbf{1},\mathbf{2,2,1}\right)\nonumber
 \\
 & = &
 \mathfrak{g}_{2}\oplus\mathfrak{su}_{2}\oplus\mathfrak{su}_{2}\oplus\left(\mathbf{7},\mathbf{2,2}\right)\oplus2\cdot\left(\mathbf{7},\mathbf{1,1}\right)\oplus\left(\mathbf{1},\mathbf{2,2}\right)\oplus3\cdot\left(\mathbf{1},\mathbf{1,1}\right)\nonumber
 \\
 & = &
 \mathfrak{g}_{2}\oplus\mathfrak{su}_{2,d}\oplus3\cdot\left(\mathbf{7},\mathbf{1}\right)\oplus4\cdot\left(\mathbf{1},\mathbf{1}\right)\oplus\left(\mathbf{7},\mathbf{3}\right)\oplus2\cdot\left(\mathbf{1},\mathbf{3}\right).
\end{eqnarray}
Therefore, by suitably branching (\ref{isom3-plane}) and (\ref{isot3-plane}),
one obtains
\begin{eqnarray}
\mathfrak{isom}\left(\mathbb{T}P_{III}^{2}\right) & \simeq &
\mathfrak{g}_{2}\oplus\mathfrak{sp}_{6}\oplus2\cdot\left(\mathbf{7},\mathbf{14}\right)\oplus\left(\mathbf{1},\mathbf{14}\right)\nonumber \\
 & = &
 \mathfrak{g}_{2}\oplus\mathfrak{su}_{2}\oplus\mathfrak{su}_{2}\oplus2\cdot\left(\mathbf{7},\mathbf{5,1}\right)\oplus2\cdot\left(\mathbf{7},\mathbf{3,3}\right)\oplus\left(\mathbf{1},\mathbf{5,3}\right)\oplus\left(\mathbf{1},\mathbf{5,1}\right)\oplus\left(\mathbf{1},\mathbf{3,3}\right)\nonumber
 \\
 & \simeq &
 \mathfrak{isom}\left(\left(\mathbb{H}\otimes\mathbb{O}\right)P^{2}\right)\oplus\left(\mathbf{7},\mathbf{5,1}\right)\oplus\left(\mathbf{7},\mathbf{3,3}\right)\oplus\left(\mathbf{1},\mathbf{5,1}\right)\oplus\left(\mathbf{1},\mathbf{3,3}\right),\\
\mathfrak{isot}\left(\mathbb{T}P_{III}^{2}\right) & \simeq &
\mathfrak{g}_{2}\oplus\mathfrak{su}_{2}\oplus6\cdot\left(\mathbf{7},\mathbf{1}\right)\oplus4\cdot\left(\mathbf{1},\mathbf{1}\right)\oplus2\cdot\left(\mathbf{7},\mathbf{3}\right)\oplus4\cdot\left(\mathbf{1},\mathbf{3}\right)\nonumber
\\
 & \simeq &
 \mathfrak{isot}\left(\left(\mathbb{H}\otimes\mathbb{O}\right)P^{2}\right)\oplus3\cdot\left(\mathbf{7},\mathbf{1}\right)\oplus\left(\mathbf{7},\mathbf{3}\right)\oplus2\cdot\left(\mathbf{1},\mathbf{3}\right),
\end{eqnarray}
and thus
\begin{eqnarray}
\mathfrak{isom}\left(\left(\mathbb{H}\otimes\mathbb{O}\right)P^{2}\right) & \subsetneq & \mathfrak{isom}\left(\mathbb{T}P_{III}^{2}\right),\\
\mathfrak{isot}\left(\left(\mathbb{H}\otimes\mathbb{O}\right)P^{2}\right) & \subsetneq & \mathfrak{isot}\left(\mathbb{T}P_{III}^{2}\right).
\end{eqnarray}
Considering (\ref{TIII-plane}) allows one to conclude that
the (tangent space) algebra of $\mathbb{T}P_{III}^{2}$ is exactly
the double of the (tangent space) algebra of $\left(\mathbb{H}\otimes\mathbb{O}\right)P^{2}$,
thus implying
\begin{align}
\mathfrak{isom}\left(\left(\mathbb{C}\otimes\mathbb{O}\right)P^{2}\right) & \subsetneq\mathfrak{isom}\left(\mathbb{T}P_{III}^{2}\right),\\
\mathfrak{isot}\left(\left(\mathbb{C}\otimes\mathbb{O}\right)P^{2}\right) & \subsetneq\mathfrak{isot}\left(\mathbb{T}P_{III}^{2}\right).
\end{align}

\begin{center}
\begin{table}
\begin{centering}
{\footnotesize{}}%
\begin{tabular}{|c|c|c|}
\hline
{\footnotesize{}Algebra} & {\footnotesize{}Definition} & {\footnotesize{}Relations with Exc. Algebras}\tabularnewline
\hline
\hline
{\footnotesize{}$\mathfrak{A}_{1}$} & {\footnotesize{}$\begin{array}{c}
\mathfrak{su}_{2}\oplus\mathfrak{f}_{4}\oplus\left(\mathbf{7},\mathbf{26}\right)\end{array}$} &
{\footnotesize{}$\mathfrak{f}_{4}\subset\mathfrak{e}_{6}\subset\mathfrak{A}_{1}\subset\mathfrak{e}_{8}$}\tabularnewline
\hline
{\footnotesize{}$\mathfrak{A}_{2}^{\text{enh}}$} &
{\footnotesize{}$\mathfrak{g}_{2}\oplus\mathfrak{su}_{2}\oplus\left(\mathbf{7},\mathbf{7+}3\cdot\mathbf{1}\right)$} &
{\footnotesize{}$\mathfrak{f}_{4}\subset\mathfrak{e}_{6}\subset\mathfrak{A}_{2}^{\text{enh}}\subset\mathfrak{e}_{8}$}\tabularnewline
\hline
{\footnotesize{}$\mathfrak{A}_{3}$} &
{\footnotesize{}$\mathfrak{g}_{2}\oplus\mathfrak{sp}_{6}\oplus\left(2\cdot\mathbf{7}+\mathbf{1},\mathbf{14}\right)$} &
{\footnotesize{}$\mathfrak{f}_{4}\subset\mathfrak{e}_{6}\subset\mathfrak{e}_{7}\subset\mathfrak{A}_{3}$}\tabularnewline
\hline
\end{tabular}{\footnotesize\par}
\par\end{centering}
\centering{}\caption{\label{tab:Relations The-three-Dixon-Rosenfeld}The three Lie algebras
of isometries corresponding to the three Dixon-Rosenfeld plane in
their relationships with the exceptional Lie algebras $\mathfrak{f}_{4},\mathfrak{e}_{6},\mathfrak{e}_{7}$
and $\mathfrak{e}_{8}$.}
\end{table}
\par\end{center}

\section{\noun{dixon-rosenfeld lines and planes}}
\label{DRplanes}

In our previous work \cite{Corr-3Rosenfeld}, we analyzed Dixon-Rosenfeld
{\rr lines} in a similar fashion, resulting in the following
definitions as homogeneous spaces
{\rr
\begin{eqnarray}
\mathbb{T}P_{I}^{1} &\simeq &\frac{SU_{2}\times SO_{9}\ltimes \left( \mathbf{%
7},\mathbf{9}\right) }{SU_{2}\times G_{2}\ltimes \left( \left( \mathbf{7},%
\mathbf{1}\right) \times 2\cdot \left( \mathbf{1},\mathbf{7}\right) \right) }%
, \\
\mathbb{T}P_{II}^{1} &\simeq &\frac{SU_{2}\times G_{2}\ltimes \left( \left(
2\cdot \mathbf{7}+\mathbf{1},\mathbf{7}\right) \times \left( \mathbf{7}+%
\mathbf{1},\mathbf{1}\right) \right) }{SU_{2}\times G_{2}\ltimes \left(
\mathbf{7},\mathbf{7}\right) }, \\
\mathbb{T}P_{III}^{1} &\simeq &\frac{SO_{5}\times G_{2}\ltimes \left(
\mathbf{5},2\cdot \mathbf{7}+\mathbf{1}\right) }{SU_{2}\times G_{2}\ltimes
\left( \left( \mathbf{1},2\cdot \mathbf{7}+\mathbf{1}\right) \times \left(
\mathbf{3},\mathbf{1}\right) \right) }.
\end{eqnarray}
}
For consistency, we now check that each Dixon-Rosenfeld line is a
strict sub-manifold of the corresponding Dixon-Rosenfeld plane. Indeed,
we notice that since
\[
\mathfrak{su}_{2}\oplus\mathfrak{so}_{9}\oplus\left(\mathbf{7},\mathbf{9}\right)\subset\mathfrak{su}_{2}\oplus\mathfrak{f}_{4}\oplus\left(\mathbf{7},\mathbf{26}\right),
\]
the isometry Lie algebra of the Dixon-Rosenfeld line $\mathbb{T}P_{I}^{1}$
is a subalgebra of $\mathbb{T}P_{I}^{2}$. For the isotropy
Lie algebra $\mathfrak{isot}\left(\mathbb{T}P_{I}^{1}\right)$, we
have
%{\rr
\begin{eqnarray}
\mathfrak{su}_{2}\oplus \mathfrak{so}_{7}\oplus \left( \left( \mathbf{7},%
\mathbf{1}\right) \oplus (\mathbf{1},\mathbf{7})\right)  &=&\mathfrak{su}%
_{2}\oplus \mathfrak{g}_{2}\oplus \left( \left( \mathbf{7},\mathbf{1}\right)
\oplus 2\cdot (\mathbf{1},\mathbf{7})\right)   \nonumber \\
&\subset &\mathfrak{su}_{2}\oplus \mathfrak{so}_{7}\oplus \left( \left(
\mathbf{7},3\cdot \mathbf{1}+\mathbf{7}\right) \oplus (\mathbf{1},2\cdot
\mathbf{7}+\mathbf{1})\right)   \nonumber \\
&=&\mathfrak{su}_{2}\oplus \mathfrak{g}_{2}\oplus \left( \left( \mathbf{7}%
,3\cdot \mathbf{1}+\mathbf{7}\right) \oplus (\mathbf{1},3\cdot \mathbf{7}+%
\mathbf{1})\right).
\end{eqnarray}
%}
It follows that the first Dixon-Rosenfeld line $\mathbb{T}P_{I}^{1}$
is a submanifold of the first Dixon-Rosenfeld plane $\mathbb{T}P_{I}^{2}$,
i.e.,
\begin{equation}
\mathbb{T}P_{I}^{1}\subsetneq\mathbb{T}P_{I}^{2}.
\end{equation}

Something similar happens for the second Dixon-Rosenfeld line $\mathbb{T}P_{II}^{1}$, since
the enhanced isometry Lie algebra is a subalgebra of the enhanced
isometry Lie algebra of the second Dixon-Rosenfeld plane $\mathbb{T}P_{II}^{2}$,
i.e.,
%{\rr
\begin{eqnarray}
\mathfrak{su}_{2}\oplus \mathfrak{so}_{7}\oplus \left( \mathbf{7+1},\mathbf{%
7+1}\right)  &=&\mathfrak{su}_{2}\oplus \mathfrak{g}_{2}\oplus \left(
\mathbf{7},\mathbf{7+1}\right) \oplus \left( \mathbf{1},2\cdot \mathbf{7+1}%
\right)   \nonumber \\
&\subset &\mathfrak{su}_{2}\oplus \mathfrak{so}_{7}\oplus \left( \left(
3\cdot \mathbf{7+}4\cdot \mathbf{1},\mathbf{7}\right) \oplus 2\cdot \left(
\mathbf{7}+\mathbf{1},\mathbf{1}\right) \right)   \nonumber \\
&=&\mathfrak{su}_{2}\oplus \mathfrak{g}_{2}\oplus \left( \left( 3\cdot
\mathbf{7+}5\cdot \mathbf{1},\mathbf{7}\right) \oplus 2\cdot \left( \mathbf{7%
}+\mathbf{1},\mathbf{1}\right) \right) .
\end{eqnarray}
%}
For the isotropy Lie algebra $\mathfrak{isot}\left(\mathbb{T}P_{II}^{1}\right)$,
we have
\begin{equation}
\mathfrak{so}_{7}\oplus\mathfrak{su}_{2}\subset\mathfrak{so}_{7}\oplus\mathfrak{su}_{2}\oplus\left(\mathbf{7},\mathbf{7+}2\cdot\mathbf{1}\right).
\end{equation}
Thus, we find that the second Dixon-Rosenfeld line $\mathbb{T}P_{II}^{1}$
is a submanifold of the second Dixon-Rosenfeld plane $\mathbb{T}P_{II}^{2}$,
i.e.,
\begin{equation}
\mathbb{T}P_{II}^{1}\subsetneq\mathbb{T}P_{II}^{2}.
\end{equation}

Finally, for the third Dixon-Rosenfeld line, we notice that
\begin{equation}
\mathfrak{g}_{2}\oplus\mathfrak{so}_{5}\oplus\left(2\cdot\mathbf{7}+\mathbf{1},\mathbf{5}\right)\subset\mathfrak{g}_{2}\oplus\mathfrak{sp}_{6}\oplus\left(2\cdot\mathbf{7}+\mathbf{1},\mathbf{14}\right).
\end{equation}
Since
\begin{align}
\notag\begin{array}{c}
\mathfrak{g}_{2}\oplus\mathfrak{su}_{2}\oplus\\
\left(2\cdot\mathbf{7}+\mathbf{1},\mathbf{1}\right)\oplus\left(\mathbf{1},\mathbf{3}\right)
\end{array}\subset\begin{array}{c}
\mathfrak{g}_{2}\oplus\mathfrak{su}_{2}\oplus\\
\left(\mathbf{7+1},4\cdot\mathbf{1}+2\cdot\mathbf{3}\right)\oplus\\
2\cdot\left(\left(\mathbf{7},\mathbf{1}\right)+\left(\mathbf{1},\mathbf{3}\right)\right),
\end{array}
\end{align}
it follows that the third Dixon-Rosenfeld line $\mathbb{T}P_{III}^{1}$
is also a submanifold of the third Dixon-Rosenfeld plane $\mathbb{T}P_{III}^{2}$,
i.e.,
\begin{equation}
\mathbb{T}P_{III}^{1}\subsetneq\mathbb{T}P_{III}^{2}.
\end{equation}

\section{\noun{conclusions and further developments }}

In this work, we introduced three inequivalent coset manifolds, i.e.,
$\mathbb{T}P_{I}^{2}$, $\mathbb{T}P_{II}^{2}$, and $\mathbb{T}P_{III}^{2}$,
that arise from the same procedure that leads to the definition of
the octonionic projective plane {\rr (also called Cayley-Moufang plane)} and of the octonionic Rosenfeld planes
as homogeneous spaces. {\rr We refer to such cosets as to the Dixon-Rosenfeld planes, since they can be regarded as generalized projective planes over the Dixon algebra $\mathbb{T}=(\mathbb{R}\otimes)\mathbb{C}\otimes\mathbb{H}\otimes\mathbb{O}$.} More specifically, we introduced three different
Lie algebras, i.e., $\mathfrak{A}_{1}$, $\mathfrak{A}_{2}^{\text{enh}}$
and $\mathfrak{A}_{3}$, arising from Tits' formula {\rr applied to the
isometry algebras of the Dixon-Rosenfeld planes. Such algebras can be regarded as analogues of the exceptional Lie algebras $\mathfrak{f}_{4},\mathfrak{e}_{6},\mathfrak{e}_{7}$ and
$\mathfrak{e}_{8}$. While the latter are isometries of the octonionic plane $\mathbb{O}P^{2}$
as well as of the octonionic Rosenfeld planes $\left(\mathbb{C}\otimes\mathbb{O}\right)P^{2}$,
$\left(\mathbb{H}\otimes\mathbb{O}\right)P^{2}$, and $\left(\mathbb{O}\otimes\mathbb{O}\right)P^{2}$, respectively, the former are isometries of the Dixon-Rosenfeld planes.}
{\rr The various relationships between these two classes of Lie algebras are summarized in Table \ref{tab:Relations The-three-Dixon-Rosenfeld}. For consistency and completeness, we also discussed the embedding of the so-called Dixon-Rosenfeld lines into the aforementioned Dixon-Rosenfeld planes.}

From the treatment given in Section 6, we obtained two different {\rr realizations} of the Dixon
algebra $\mathbb{T}$ as the $\left(\boldsymbol{7}+\boldsymbol{1},\boldsymbol{7}+\boldsymbol{1}\right)$
and as the $2\cdot\left(\mathbf{7}+\mathbf{1},\mathbf{3}+\mathbf{1}\right)$
of $\mathfrak{g}_{2}\oplus\mathfrak{su}_{2}$ which might be interesting
for physical applications
%{\rr
The two realizations can be related by trading the spin-$3$ representation of $\mathfrak{su}_{2}$ for a pair of spin-$1$ (vector) representations and a spin-$0$ (scalar) representation. In future work, it will be worthwhile to investigate the potential relationship of the isometry Lie algebras/groups of the Dixon-Rosenfeld planes with an interesting
model for three generations of fermions in the Standard Model of particle physics recently introduced by Furey and Hughes \cite{Furey:2024yzz}.
%}
%In a forthcoming paper, we will
%analyze similar constructions for the Dixon-Rosenfeld lines obtaining
%coherent representations for $\mathbb{T}$ and suitable embeddings
%of the Dixon-Rosenfeld lines into the Dixon-Rosenfeld planes.

\section{\noun{acknowledgments} }

%The work of D.Corradetti is supported by a grant of the Quantum Gravity
%Research Institute. The work of AM is supported by a ``Maria Zambrano''
%distinguished researcher fellowship, financed by the European Union
%within the NextGenerationEU program.

Thank you to Cohl Furey for discussions relating to the Dixon algebra and her work with Mia Hughes on three generations of fermions in the standard model.

\end{document}